%% file: main.tex
\documentclass[journal]{vgtc}                     

\onlineid{0}

\vgtccategory{Research}

\title{Interactive Mascot: A Scene-Centric\\Interaction Grammar for Data Visualizations}

\author{%
  \authororcid{Zhicheng Liu}{0000-0002-1015-2759}
}

\authorfooter{
  \item
  	Zhicheng Liu is with University of Maryland College Park.
  	E-mail: leozcliu@umd.edu
}

\input{tex/abstract}
\keywords{Interaction, data visualization, grammar}

\teaser{
  \centering
  \includegraphics[width=\linewidth, alt={A view of clouds with orange sunrays shining through from behind.}]{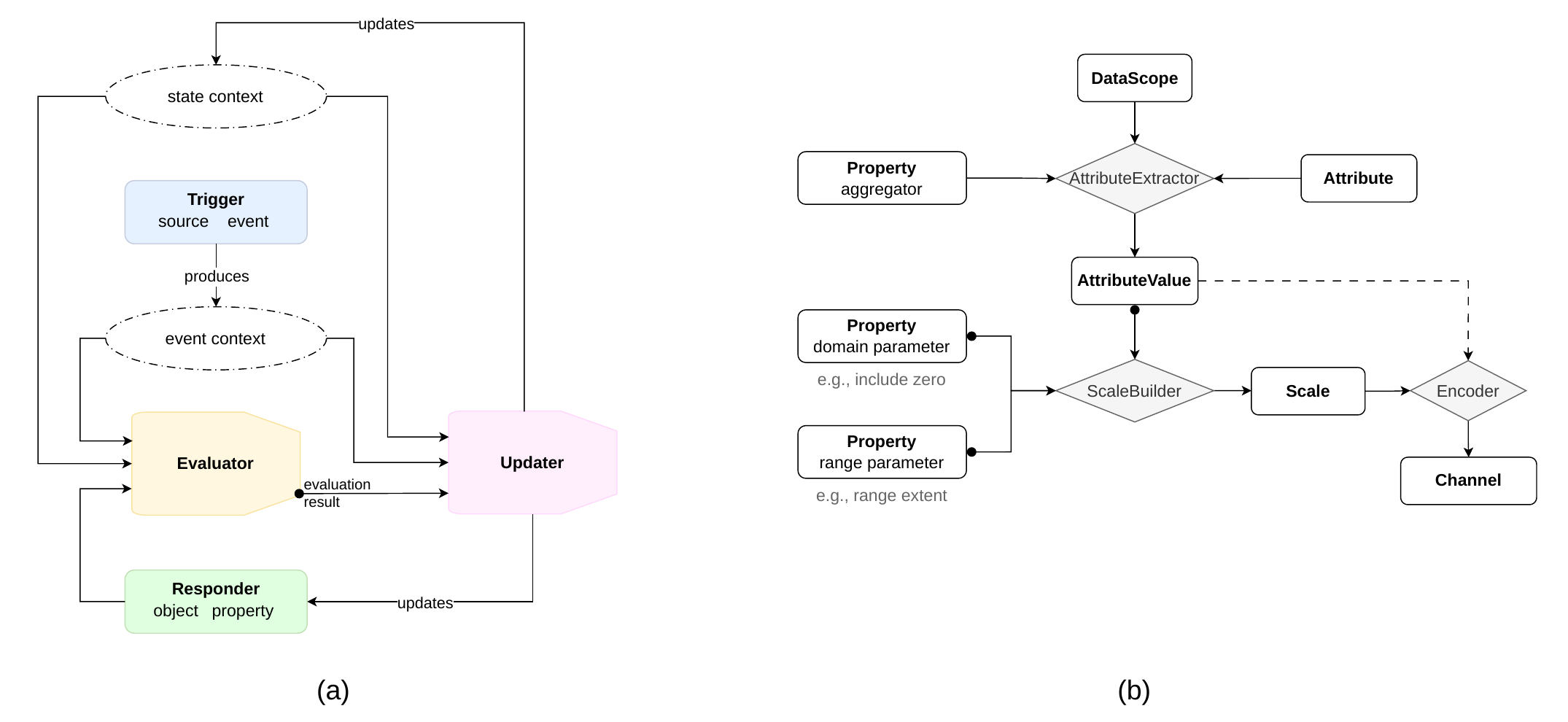}
  \caption{%
  	(a) Interactive Mascot identifies four core interaction components---trigger, responder, evaluator, and updater---for users to specify data visualization interactivity. The grammar also defines the permissible information-flow relationships among these four components and two forms of context (event context and state context). State variables maintained in the state context may serve as trigger sources. Stateful interactions may also update state variables, represented as properties of the state context. In the diagram, unlabeled edges indicate that the origin node serves as an input to the destination node. (b) To realize the high-level interaction semantics, \name introduces a dependency-graph execution model based on a set of reusable dependency graph patterns. Each pattern corresponds to a semantic visualization component and captures its dependency structure.
    The dependency graph shown here illustrates a reusable dependency graph pattern corresponding to a visual encoding component. 
  }
  \label{fig:teaser}
}

\graphicspath{{figs/}{figures/}{pictures/}{images/}{./}} 

\usepackage{graphicx}
\usepackage{xspace}
\usepackage{float}
\usepackage{placeins}
\usepackage{tikz}         
\usepackage{xcolor}       
\usepackage{booktabs}
\usepackage{multirow}
\usepackage{makecell,array}
\usepackage{colortbl}
\usepackage{graphicx}
\usepackage{listings}
\usepackage{mdframed}
\usepackage{dblfloatfix}
\usepackage{cite}
\usepackage[scaled=0.95]{inconsolata}
\usepackage[ruled,vlined]{algorithm2e}
\usepackage{url}
\usepackage{svg}

\input{tex/commands}

\begin{document}



\maketitle


\input{tex/1.intro}
\input{tex/2.methods}

\input{tex/2.components}
\input{tex/3.depGraph}
\input{tex/4.mascot}

\input{tex/5.related_work}
\input{tex/6.evaluation}

\input{tex/7.conclusion}


\bibliographystyle{abbrv-doi-hyperref}

\bibliography{references}

\input{tex/appendix.tex}

\end{document}

%% file: tex/abstract.tex
\abstract{Scene-centric visualization systems expose semantic components, such as marks, encodings, layouts, and axes, as first-class objects that can be directly manipulated. Existing interaction abstractions, however, are largely based on event streams, signals, and data selections rather than semantic scene components. This mismatch makes interactions involving scene components less natural to specify and limits the expressive power of scene-centric visualization systems. We present Interactive Mascot, a scene-centric interaction grammar for data visualizations. Interactive Mascot extends scene-centric representations for static visualizations by modeling interactive behavior as information flow among four interaction components (trigger, responder, evaluator, and updater) and two forms of context (event context and state context). 
To realize these semantics, we introduce a dependency-graph execution model that systematically transforms interaction specifications into executable dependency graphs using reusable graph patterns associated with semantic visualization components.
We implement Interactive Mascot in the JavaScript library Mascot.js and evaluate its performance and usability. Interactive Mascot 
achieves runtime performance comparable to Vega-Lite, and a qualitative user study shows that the grammar is learnable and usable for interaction authoring.}

%% file: tex/commands.tex
\definecolor{coolblack}{rgb}{0.0, 0.18, 0.39}
\definecolor{markcolor}{RGB}{197, 90, 17}
\definecolor{refmarkcolor}{RGB}{103, 103, 103}
\definecolor{groupcolor}{RGB}{143, 39, 38}
\definecolor{layoutcolor}{RGB}{121, 61, 134}
\definecolor{datascopecolor}{RGB}{84, 130, 53}
\definecolor{enccolor}{RGB}{68, 114, 196}
\definecolor{constrcolor}{RGB}{216, 83, 214}
\definecolor{viewcolor}{RGB}{211, 171, 44}
\definecolor{lightgray}{RGB}{212,212,212}
\definecolor{lightestgray}{RGB}{245,245,245}
\definecolor{darkgreen}{RGB}{0, 153, 51}
\definecolor{eeeColor}{RGB}{246, 246, 246}
\definecolor{kwblue}{HTML}{4682B4}
\definecolor{varTeal}{HTML}{007C7C}
\definecolor{strgreen}{HTML}{85A16A}
\definecolor{numPurple}{HTML}{6A0DAD}
\definecolor{orange}{HTML}{8A3324}

\newcommand{\eg}{{e.g.,}\xspace}
\newcommand{\ie}{{i.e.,}\xspace}
\newcommand{\etal}{{et al.}\xspace}

\newcommand{\bpstart}[1]{\par\smallskip \noindent{\textbf{#1.}}}

\newcommand{\variable}[1]{\tikz[baseline=(X.base)] \node[draw=lightgray, rounded corners=2pt, rectangle, inner sep=1.5pt] (X) {\small #1};}

\newcommand{\operator}[1]{{\small $<$#1$>$}}

\newcommand{\mk}[1]{\textcolor{markcolor}{#1}}
\newcommand{\group}[1]{\textcolor{groupcolor}{#1}}
\newcommand{\layout}[1]{\textcolor{layoutcolor}{#1}}
\newcommand{\datascope}[1]{\textcolor{datascopecolor}{#1}}
\newcommand{\enc}[1]{\textcolor{enccolor}{#1}}
\newcommand{\constr}[1]{\textcolor{constrcolor}{#1}}
\newcommand{\refmark}[1]{\textcolor{refmarkcolor}{#1}}

\newcommand{\attr}[1]{{\scriptsize $\mathsf{#1}$}}

\newcommand{\inlineCode}[1]{{\small $\texttt{#1}$}}

\newenvironment{supp}
  {\begin{mdframed}[
    backgroundcolor=eeeColor,
    font=\small,
    linewidth=0pt,
    innertopmargin=5pt,
    skipabove=3pt,
    innerleftmargin=4pt,
    innerrightmargin=1pt
  ]}
  {\end{mdframed}}

\newcommand{\name}{Interactive Mascot\xspace}

\newcommand{\source}{\ensuremath{\sigma}}
\newcommand{\event}{\ensuremath{e}}
\newcommand{\context}{\ensuremath{c_{e}}} 
\newcommand{\stCtx}{\ensuremath{c_{s}}} 

\newcommand{\component}{\ensuremath{\kappa}\xspace}
\newcommand{\property}{\ensuremath{p}}
\newcommand{\updater}{\ensuremath{\mathcal{U}}}

\newcommand{\evaluator}{\ensuremath{\mathcal{E}}}

\lstdefinestyle{js} {%
  morekeywords={[3]{scn, dt, circle, collection, xEnc, trigger1, responder1, updater1, trigger2, responder2, evaluator, updater2, xAxis}},
  morekeywords={[2]{scene, import, mark, repeat, divide, encode, find, align, affix, legend, axis, gridlines, activate}},
  morekeywords={[1]{typeof, new, catch, function, return, null, catch, switch, var, if, in, while, do, else, case, break, let}},
  basicstyle=\footnotesize\ttfamily, 
  abovecaptionskip=4pt,
  belowcaptionskip=6pt,
  aboveskip=6pt,
  belowskip=6pt,
  columns=flexible,
  frame=tb,
  numbers=left,
  stepnumber=1,
  firstnumber=1,
  numbersep=3pt,
  numberfirstline=true,
  numberstyle=\color{coolblack},
  identifierstyle=\color{black},
  keywordstyle={[1]{\color{coolblack}\bfseries}},
  keywordstyle={[2]{\color{coolblack}\bfseries}},
  keywordstyle={[3]{\color{coolblack}}},
  stringstyle=\color{coolblack}\ttfamily,
  morestring=[b]",
  morestring=[b]',
  alsodigit={:;},	
  tabsize=2,
  showtabs=false,
  showspaces=false,
  showstringspaces=false,
  extendedchars=true,
  breaklines=true,
  rulecolor=\color{gray},
}

%% file: tex/1.intro.tex
\section{Introduction}
Many approaches to data visualization creation 
\cite{liu_atlas_2021,liu_manipulable_2025, ren_charticulator_2018,liu_data_2018,tsandilas_structgraphics_2020, cui_mixed-initiative_2022, chen_mystique_2023, chen_towards_2020,wang_animated_2021,shi_piccl_2026} adopt a scene-centric representation:
semantic visualization components are represented as first-class objects that users can directly instantiate, 
query, reference, modify, and organize  
through an API (application programming interface) or a GUI (graphical user interface). 
Unlike declarative grammars where marks and encodings appear only as specification constructs, scene-centric systems expose the visualization scene itself as a manipulable object model. For example, Mascot.js \cite{liu_manipulable_2025} presents an object model to describe semantic components such as mark, collection, and encoding, and formulates generative and modificative procedures to create and modify these components; similarly, PiCCL \cite{shi_piccl_2026} organizes visual objects into a hierarchical object tree, and provides composition operations over these objects. 
Such scene-centric representations  allow users to reason about visualization structure at the level of semantic scene components rather than low-level graphical primitives 
\cite{xia_dataink_2018,liu_data_2018,shi_piccl_2026,tsandilas_structgraphics_2020},  exhibit considerable expressive power beyond standard charts to support bespoke designs \cite{liu_manipulable_2025,shi_piccl_2026,pollock_gofish_2026,liu_data_2018}, and demonstrate potential to support downstream applications in visualization analysis, augmentation and reuse \cite{chen_mystique_2023,chen_towards_2020,cui_mixed-initiative_2022,wang_animated_2021}. However, these formulations focus primarily on static representations of data, and do not  support composing interactivity in the visualizations they produce.

Existing abstractions of interactions in visualization, on the other hand, are largely designed around event streams, data selections, signals, or interactors rather than semantic scene components. Reactive Vega \cite{satyanarayan_reactive_2016} describes interaction through low-level primitives such as signals and event streams; Vega-Lite \cite{satyanarayan_vega-lite_2016} provides higher-level abstractions through selection components such as predicate, transforms, and resolve. These abstractions encapsulate the scene behind declarative specifications, and do not treat scene components as first-class citizens. Libra.js \cite{zhao_libra_2025} 
focuses on modular interaction services. 
These approaches do not operate directly on semantic scene components.
Liu et al. \cite{liu_spatial_2024} introduce a model to directly manipulate visual objects in a static visualization based on spatial constraints such as gravity, support and collision. This model takes a scene-centric perspective and aligns closely with the goal of this paper. However, it exhibits considerably less expressive power compared to the aforementioned approaches.  

This mismatch is problematic for scene-centric visualization systems. Many interactions that are naturally defined in terms of scene components (e.g., dragging an axis to modify its scale range, manipulating layouts, or coordinating scene objects across multiple events) must be reformulated in terms of data selections, event streams, or low-level signals. This translation obscures the original interaction intent, makes specifications less natural to articulate, and leaves certain interaction techniques  difficult to express. We therefore argue that
a unified abstraction is necessary in scene-centric systems to specify interactions in terms of the same semantic components used to construct visualizations, rather than through a separate interaction vocabulary.

To address this limitation, we propose \name, a scene-centric interaction grammar that extends scene-centric representations to support interactive behavior. 
Interactive Mascot makes two complementary contributions. First, it models interactive behavior as information flow among \textit{four interaction components} (trigger, responder, evaluator, and updater) and  \textit{two forms of context} (event context and state context), enabling both transient and persistent interaction behaviors. Programmers specify interactions using the four components, while the grammar defines the permissible information-flow relationships among the components and contexts. Second, 
to realize these high-level interaction semantics, Interactive Mascot 
introduces a dependency-graph execution model based on a set of reusable dependency graph patterns, each corresponding to a semantic visualization component.
These patterns serve as templates to systematically  instantiate interaction specifications as dependency graphs, enabling interaction execution through the propagation of changes along dependency paths.  This design separates interaction semantics from interaction execution, 
allowing programmers to specify interactions at the level of semantic components,  while the underlying dependency graph automatically coordinates the low-level execution required to maintain consistency among visualization components.

We present technical details on the implementation of \name in the JavaScript library Mascot.js, and evaluate the grammar in terms of performance and usability. 
\name 
achieves comparable runtime performance to Vega-Lite in frame rate benchmarks.
A qualitative study on interaction authoring tasks demonstrates that the grammar is usable and learnable. 

%% file: tex/2.methods.tex
\section{Design Goals and Methodology}
In designing \name, we focus on the following three desired properties of the grammar:

\begin{enumerate}{}{}
\item{\textbf{Scene-centric}: interactions should be specified using the same semantic scene components that were used to construct static visualizations.} 
\item{\textbf{High-level}: the grammar should expose abstractions focusing on user intent rather than execution details. Low-level mechanisms such as event handling, update ordering, and dependency maintenance should be hidden from users and automatically handled by the underlying execution model.}
\item{\textbf{Expressive}: the grammar should support a wide range of interaction designs found in practice, including both conventional data exploration techniques and direct manipulations of visualization components.}
\end{enumerate}

To realize these goals, we conducted an iterative process consisting of interaction analysis, grammar design, execution-model design, and implementation. We build upon prior work on visualization languages \cite{liu_manipulable_2025,shi_piccl_2026}, authoring systems \cite{snyder_divi_2024,liu_data_2018,ren_charticulator_2018}, and theoretical frameworks \cite{satyanarayan_critical_2019} that propose component-based representations of static visualizations. Through a review of these works, we identified a set of core scene components that recur across existing approaches (Sec. \ref{sec:static-compnts}). These scene components represent semantic objects that can participate in interaction specification. 

We then analyzed representative interactive visualizations from the online galleries of D3 and Vega/Vega-Lite, spanning diverse user intents \cite{yi_toward_2007}. We selected these libraries because they collectively cover a broad spectrum of interaction designs commonly used in interactive visualizations, ranging from low-level custom interactions to higher-level declarative interactions. The examples encompass diverse interaction techniques, including selection, filtering, navigation, brushing and linking, and coordinated views. Together, they establish the target level of expressiveness for the grammar and serve as the basis for our analysis.

Our analysis was conducted at two levels. At the semantic level, we sought recurring semantic roles shared by diverse interaction techniques. Specifically, we examined: (1) what scene components participate in an interactive behavior, (2) what semantic roles these components assume, and (3) what information flows connect these roles and components in a cause-and-effect relationship. We sought abstractions that provide succinct and uniform vocabularies for answering these questions across a wide range of interaction designs.

At the execution level, our goal was to identify a generic mechanism for coordinating updates among scene components. Inspired by the success of graph-based approaches in UI toolkits \cite{vander_zanden_lessons_2001,myers_amulet_1997,myers_garnet_1990} and visualization systems such as Vega \cite{satyanarayan_reactive_2016}, we explored dependency graphs as a candidate execution model. Existing graph-based approaches, however, are typically designed either for GUI interaction behaviors or for dataflow-based architectures. 
The challenge in our setting was therefore to identify recurring graph structures suitable for scene-centric visualization components. 

The semantic abstraction and execution model were developed jointly. We started with a few examples and iterated between semantic-level analysis and execution-level design, refining one in response to constraints imposed by the other. Each candidate abstraction was required to admit a consistent graph-based realization across the interaction examples. As additional examples were analyzed, recurring semantic and execution structures emerged. During refinement, abstractions that exposed execution details or required interaction-specific graph structures were discarded in favor of abstractions that could be expressed uniformly across diverse examples.

Alongside the analysis, we prototyped implementations in Mascot.js \cite{liu_manipulable_2025}, which provides implementations and APIs for scene-centric static visualization components. We took care to ensure that the resulting grammar was not tied to implementation-specific details of Mascot.js, but instead captured conceptual abstractions and execution models that could be realized in other systems.

Through iterative refinement, we evaluated the evolving grammar against an expanding collection of interaction examples. Refinement ceased when additional examples no longer introduced new interaction components, information-flow patterns, or dependency-graph structures. 
Across the analyzed interactions, we consistently observed four recurring semantic roles: \textit{what initiates an interaction}, \textit{what scene component is affected}, \textit{under what conditions the effect should occur}, and \textit{how the affected component should change}. These recurring roles became the four interaction components of Interactive Mascot: \textit{trigger}, \textit{responder}, \textit{evaluator}, and \textit{updater}. The recurring information-flow relationships among them formed the basis of the interaction grammar (Sec. \ref{sec:interactive_components}). In parallel, the execution-level analysis resulted in the dependency-graph execution model (Sec. \ref{sec:patterns}). Based on these findings, we finalized the API and dependency-graph implementation in Mascot.js (Sec. \ref{sec:implementation}). 


%% file: tex/2.components.tex
\section{\name: A Scene-Centric\\Interaction Grammar}
\label{sec:abstraction}
\subsection{Scene Components in Static Visualizations}
\label{sec:static-compnts}

\begin{figure}[h]
 \centering \includegraphics[width=\linewidth]{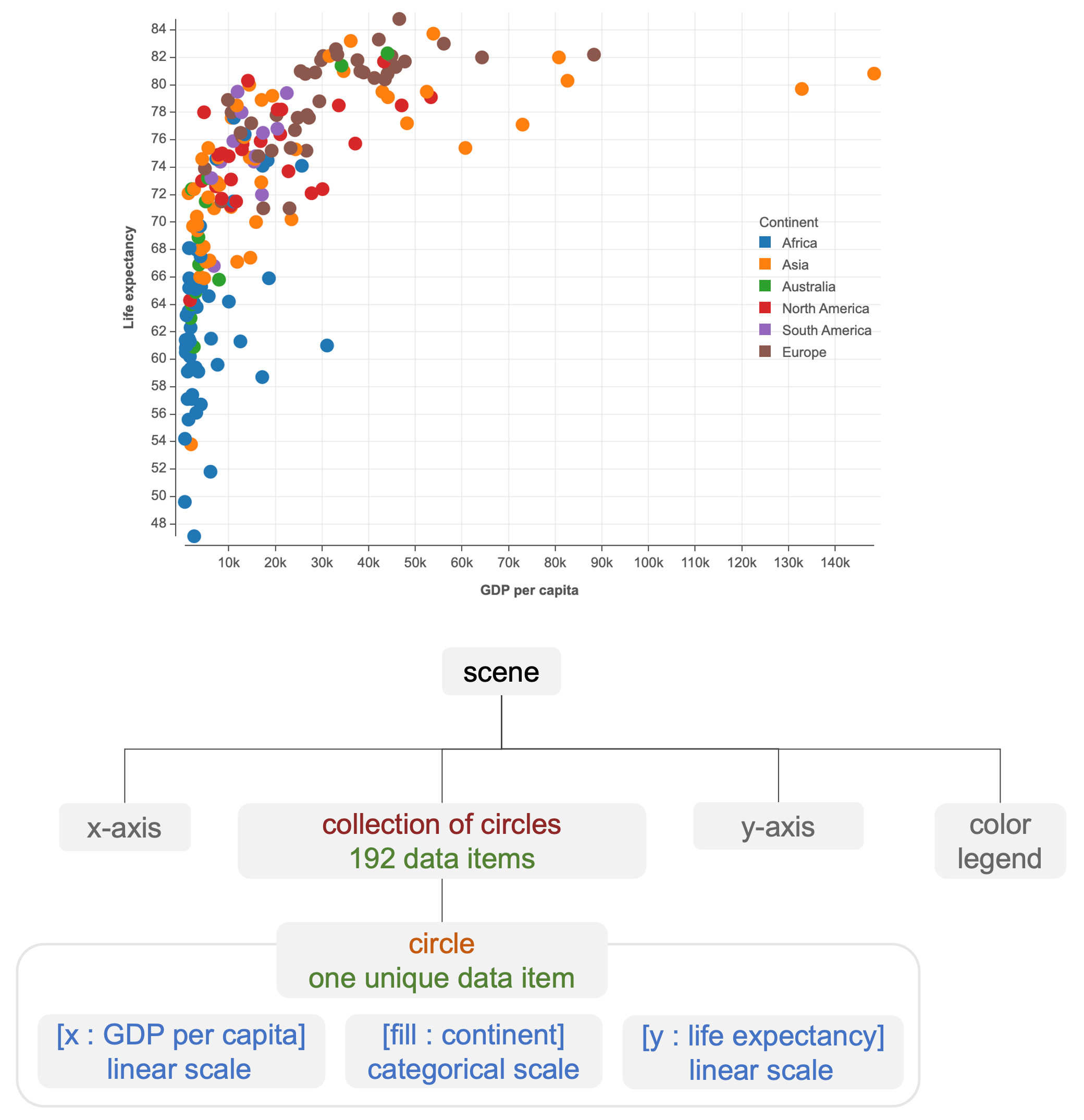}
 \caption{Semantic components in a static scatterplot. \label{fig:scatterplot}}
 \vspace{-2em}
\end{figure}

Our review of prior work \cite{satyanarayan_critical_2019,liu_manipulable_2025,shi_piccl_2026,liu_spatial_2024,snyder_divi_2024} identifies a set of core components that serve as the building blocks of static visualizations: mark, collection, data scope, encoding, layout, constraint, and reference element. Table \ref{tbl:components} provides a definition and examples for each component. To understand how some of these components are used in a static visualization, consider a scatterplot showing life expectancy and GDP per capita data for 192 countries (Figure \ref{fig:scatterplot}). The scene consists of a {\group{collection}} of \mk{circles} and five {\refmark{reference elements}} ({\refmark{x-axis}}, {\refmark{y-axis}}, \refmark{x grid lines}, \refmark{y grid lines}, and {\refmark{color legend}}). The {\datascope{data scope}} of the \group{collection} is the set of 192 data items in the dataset. The children of the \group{collection} are 192 {\mk{circle marks}}, each representing a unique data item. Three {\enc{visual encodings}} for the circle marks are defined: {\enc{x  encoding}} (\attr{GDP/PC} mapped to x~position), {\enc{y encoding}} (\attr{life~expectancy} mapped to y~position), and {\enc{fill color encoding}}  (\attr{continent} mapped to fill~color).

\begin{table}[th]
\renewcommand{\arraystretch}{1.28}
    \centering
    \begin{tabular}{p{1.45cm} p{3.5cm} p{2.6cm} p{1cm}}
    \toprule
     Component    &  Definition & Examples  \\
     \midrule
      \mk{mark}  & basic graphical element used to represent data \cite{liu_manipulable_2025,snyder_divi_2024,liu_spatial_2024,satyanarayan_critical_2019,shi_piccl_2026} & line, rectangle, text  \\
      \group{collection} & group of marks, where each mark has a unique data scope \cite{liu_manipulable_2025,satyanarayan_critical_2019,liu_data_2018,shi_piccl_2026} & bars in a bar chart, arcs in a donut chart   \\
      \datascope{data scope} & data item(s) represented by a mark or a collection \cite{liu_manipulable_2025,satyanarayan_critical_2019} &  \attr{(France, Europe, 38K, 82)} represented by a circle   \\
      \enc{encoding} & mapping of a data attribute to a visual channel of a mark \cite{liu_manipulable_2025,satyanarayan_critical_2019} & \attr{continent} mapped to fill color\\
      \layout{layout}  & algorithmic spatial arrangement of marks \cite{liu_manipulable_2025,satyanarayan_critical_2019,shi_piccl_2026} & stacking, packing, grid, force-directed\\
      \constr{constraint}  & inter-element spatial relationship \cite{liu_manipulable_2025,liu_spatial_2024,shi_piccl_2026} & alignment, affixation \\
      \refmark{reference element}  & element to facilitate interpretation \cite{liu_manipulable_2025,snyder_divi_2024,satyanarayan_critical_2019} & axis, legend, gridlines\\
     \bottomrule
    \end{tabular}
    \caption{Core components in static visualizations, synthesized from previous work \cite{snyder_divi_2024,liu_manipulable_2025,satyanarayan_critical_2019,liu_spatial_2024,shi_piccl_2026}.\label{tbl:components}}
    \vspace{-1em}
\end{table}

\subsection{Interactivity Components: Trigger and Responder}
\label{sec:interactive_components}
Based on the semantic roles identified through the analysis in Section II, Interactive Mascot introduces four interaction components:  
\textit{trigger}, \textit{responder}, \textit{evaluator}, and \textit{updater}.  A single scene component may assume different roles in different interactions. For example, a circle mark may act as a trigger when clicked, as a responder when highlighted, or as both simultaneously. Collectively, the four components answer four fundamental questions underlying an interactive behavior: what initiates it (trigger), what is affected (responder), under what conditions (evaluator), and how it changes (updater). \name also introduces two information contexts: \textit{event context} and \textit{state context}. Figure \ref{fig:teaser}a provides an overview of the information flow between these components and contexts. To illustrate these concepts, we introduce two example interactions performed on the scatterplot in Figure \ref{fig:scatterplot}:\\
\noindent{}\textbf{E1: configure scale range.} Increase or decrease the range extent of the x-axis by dragging the axis; the positions of the circles, the x-axis ticks and labels, and the vertical gridlines update accordingly. 

\noindent{}\textbf{E2: select by attribute value.} Clicking a circle will select all the countries from the same continent as the clicked circle (e.g., clicking any blue circle will select all circles representing African countries); the circles representing African countries remain unchanged while the other circles fade to gray. 

We define a \textbf{trigger} as an event source that initiates an interactive behavior in a visualization. In \name, a trigger consists of a source\footnote{We use the term source rather than target, as used in D3's event model, to denote the semantic object from which an interaction event originates (e.g., a visualization element, data scope, encoding, state variable). In contrast, target in D3 specifically refers to the DOM element that receives an event.} \source~and an event \event.

\begin{supp}
$\mathsf{trigger}:= \langle\source, \event\rangle$
\end{supp}

\noindent{A} source \source~is a semantic object capable of serving as the origin of an event \event. Examples include visualization elements (\eg mark, axis, background), UI widgets (\eg button, drop-down menu, slider), and state variables. For example, in E1, the trigger consists of the drag \textit{event} and the \refmark{x-axis} as the \textit{source}; In E2, the trigger consists of the click \textit{event} and any \mk{circle} as the \textit{source}. In more complex interactions that involve state management, a source can also be a state variable. We show in Section \ref{sec:updater_evaluator} how state variables themselves can serve as event sources in stateful interactions.

We define a \textbf{responder} as an object whose property or value is modified in response to a trigger event. In \name, a responder specification consists of an object \component~and a property \property:

\begin{supp}
    $\mathsf{responder}$ := $\langle$\component, \property $\rangle$
\end{supp}


\noindent{A} responder object \component~is usually one of the scene components defined in Section \ref{sec:static-compnts}. For instance, in E1, the responder object is the \enc{x encoding}, and the property is range extent; the responder object is different from the trigger source. In E2, every \mk{circle mark} is a responder object, and the property is the fill color; the responder object is the same as the trigger source. We use the notation \component.\property~to denote the property value (e.g., $\mathsf{circle.fill}$). In more complex cases that involve state management, the responder object may also be the state context, whose properties correspond to individual state variables  
(Section 
\ref{sec:updater_evaluator}). This differs from the trigger definition, where an individual state variable may serve as the event source. The distinction reflects the two roles of state variables in interaction: a change to a state variable can generate an event, whereas updating persistent interaction state modifies a property of the state context.

\subsection{Information Contexts: Event Context and State Context}
\label{sec: contexts}
\begin{figure}[hb]
 \centering \includegraphics[width=\linewidth]{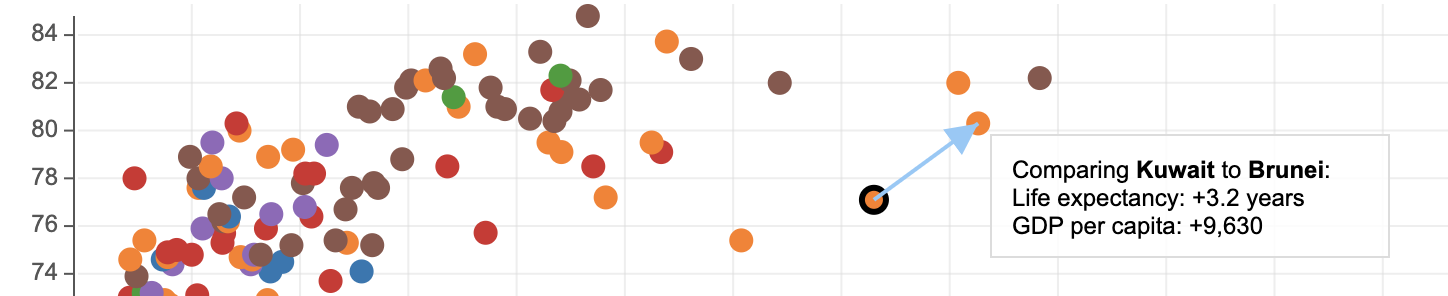}
 \caption{Comparing the hovered countries to a pinned baseline country (highlighted in black border). \label{fig:pin+compare}}
\end{figure}

\name defines two information contexts. The \textbf{event context} captures transient information associated with a triggering event, while the \textbf{state context} maintains persistent interaction state across multiple events. The event context \context~of a triggering event \event~records a set of relevant information necessary to interpret and process that event. For example, in E1, the event context captures the distance moved by the mouse cursor; in E2, the event context records the specific \mk{circle} being clicked. An event context typically captures multiple context fields depending on the type of event. Appendix A includes examples of common events and the corresponding event context fields. We use the notation \context[$f$] to denote the value of a particular field $f$ captured in the context \context~(\eg \context[``x'']). A state context \stCtx~is necessary for interactions that require information to persist across multiple events. Consider the following interaction design:

\noindent{}\textbf{E3: pin baseline then compare.} Clicking a circle pins it as a baseline for comparison, and the pinned country is highlighted by a black border; hovering over any other circle shows a blue arrow from the pinned circle to the hovered circle and a tooltip displaying the differences between the two countries in terms of the attributes (Figure \ref{fig:pin+compare}).

In this example, the event context associated with the click event identifies the pinned country. However, because event contexts are transient and do not persist beyond the current event, the pinned country must be recorded in the state context when the click occurs. This allows subsequent hover events to access the stored information. We similarly use \stCtx[$v$] to denote the value of state variable $v$ maintained in the state context (e.g., \stCtx[``pinned'']).

\subsection{Interaction Components: Updater and Evaluator}
\label{sec:updater_evaluator}
A trigger can directly cause an updater to modify the property of a responder object, where  an \textbf{updater function}  \updater~modifies the responder property based on the information in the event context and/or state context: 

\begin{supp}
\component.\property~= \updater(\component, \context, \stCtx) $\mid$ $\mathsf{trigger}:= \langle$\source, \event $\rangle$ \& $\mathsf{responder}:= \langle$\component, \property $\rangle$
\end{supp}

\noindent{In} E1, the changes in distance (\context["dx"]) moved by dragging (\event) the \refmark{x-axis} (\source) are translated into changes in the range extent (\property) of \enc{x encoding} (\component) using this function: 
{\small \updater($\mathsf{enc_x}$, \context, \stCtx) $\Rightarrow$ $\mathsf{enc_x}$.$\mathsf{range\_extent}$ + \context["dx"]}
. We call such an interactive behavior a \textbf{direct property modification}. In transient interactions such as E1, the state context is not used.

In other interaction designs, updating responders' properties is less straightforward. The responder objects are first evaluated by an \textbf{evaluator function}  \evaluator~against predefined conditions that involve the event context and/or state context, 
and we denote the outcome of the evaluation for each component \component as \evaluator(\component). 
The value of the property is then determined by an \textbf{updater function} \updater, which takes the evaluation outcome as an additional argument: 

\begin{supp}

\component.\property~= \updater(\component, \context, \stCtx, \evaluator(\component)) $\mid$ $\mathsf{trigger}:= \langle$\source, \event $\rangle$ \& $\mathsf{responder}:= \langle$\component, \property $\rangle$     

\end{supp}

\noindent{For} instance, in E2, when the user clicks (\event) to select any circle \mk{mark} (\source), the fill color (\property) of each \mk{mark} (\component) is updated based on whether it represents a country with the same data attribute \attr{Continent} as the selected mark (\context["element"]). In this case, we have the following evaluator function definition:

{\scriptsize
\noindent\evaluator(mark, \context, \stCtx)$\Rightarrow$mark.dataScope["continent"]==\context["element"].dataScope["continent"]
}
\noindent{and} the following updater function definition: 

{\scriptsize
\noindent\updater(mark, \context, \stCtx, \evaluator(mark)) $\Rightarrow$ 
if not  \evaluator(mark), then mark.fill = "gray"
}


We call such an interactive behavior a \textbf{conditional encoding}. Conditional encodings are often used in conjunction with a regular \enc{visual encoding} to determine the final property value. For example, the fill color of marks that share the same \attr{Continent} value as the selection is determined by the original \enc{fill encoding}; when the selection is cleared, each mark reverts to the fill color encoding \attr{Continent}. 

Stateful interactions often involve both direct property modifications and conditional encodings. Using E3 as an example, we define the following interactive behaviors:
\begin{enumerate}
    \item Clicking (\event) a circle \mk{mark} (\source) will perform a direct property modification on the state context (\component), where the property  ``pinned'' (\property) is updated with the clicked circle mark.
    \item A change (\event) in the state variable ``pinned'' (\source) will then trigger a conditional encoding, where an evaluator will check each circle \mk{mark} to see if the mark is the pinned state variable. The stroke width (\property) of the circles (\component) will then be updated accordingly.
    \item Subsequent hovering (\event) over a circle \mk{mark} (\source) will also involve conditional encodings to update the position and visibility of the arrow (\component) and the tooltip (\component).
\end{enumerate}
 

%% file: tex/3.depGraph.tex
\section{Execution Model}
\label{sec:patterns}
\begin{figure}[h]
 \centering \includegraphics[width=\linewidth]{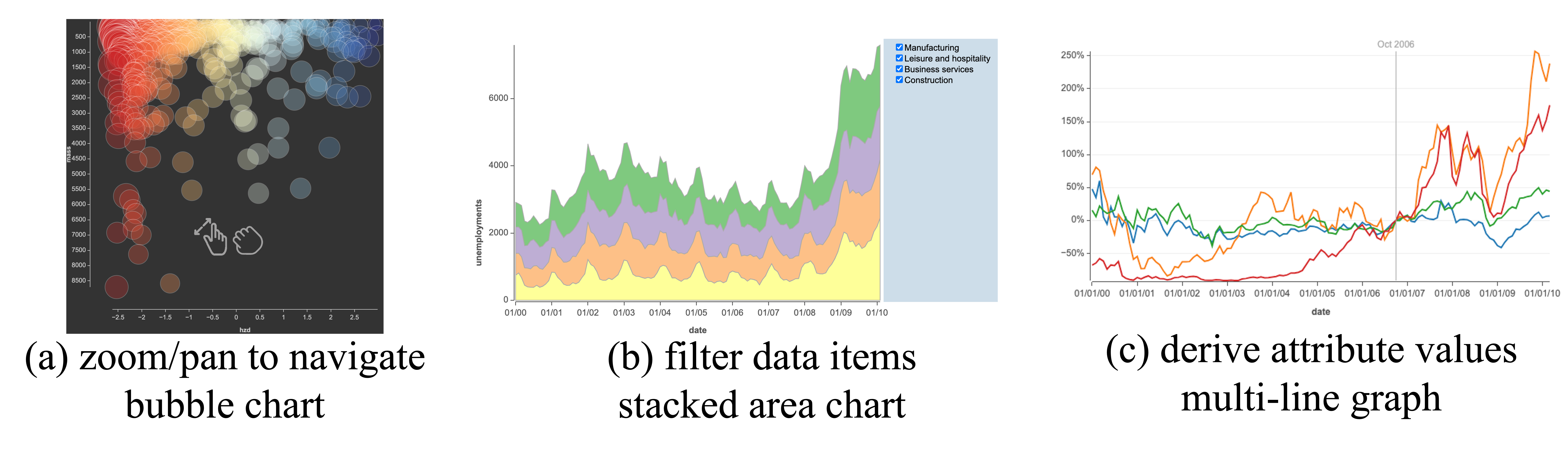}
 \caption{Modifying a component requires updating all the other relevant components to preserve chart integrity. \label{fig:interconnected}}
\end{figure}

The interaction grammar identifies \textit{what} components and contexts are necessary for interaction specification. To realize these interaction semantics, we need a robust mechanism that defines \textit{how} those semantics are realized efficiently and consistently.
A major challenge here is to maintain the integrity of data visualization: the scene components are intricately interconnected, and modifying one component can have unintended consequences that impact other components. Consider the following examples: (1) zooming and panning in a bubble chart (Figure \ref{fig:interconnected}a) -- simply modifying the scales for x- and y-position encodings is insufficient, as the corresponding axes also need to be updated in sync with the scales; (2) dynamic filtering of data items through checkboxes in a stacked area chart (Figure \ref{fig:interconnected}b) -- hiding area marks alone will create unintended gaps in the chart, necessitating the recomputation of the area marks' positions; (3) hovering over the time axis of an index chart to choose a baseline date, from which relative changes in stock prices are computed  (Figure \ref{fig:interconnected}c) -- the data values, y-position scale, mark positions, y-axis, and the baseline date annotations all need to be updated in a particular order. In each of these examples, it is important to carefully identify and update all relevant components to maintain the integrity of the chart. Given the vast design space of static charts and interaction techniques, addressing this on a case-by-case basis can be ad hoc and error prone.  

To systematically address this problem, \name uses dependency graphs as the execution model for the interaction grammar introduced in Section \ref{sec:abstraction}. The dependency graphs realize these semantics through modular operators and change propagation. The graph nodes are either \variable{variables} (representing component properties) or \operator{operators} (representing modular processors). Rather than constructing a dependency graph individually for every visualization, Interactive Mascot associates reusable graph patterns with semantic visualization components. These patterns serve as templates that are systematically instantiated to produce an executable dependency graph.
Interaction execution is realized as the propagation of changes along the dependency graph paths, where the \operator{operator} nodes on the paths are invoked and the \variable{variable} nodes' values re-evaluated. 

To provide a detailed explanation of this mechanism in \name, we start by introducing simple dependency graph patterns for some common scene components; we then illustrate the construction of a dependency graph for the scatterplot examples based on these patterns; finally, we present additional patterns to demonstrate how the architecture generalizes beyond the scatterplot examples to other types of components.

\subsection{Dependency Graph Patterns: Bounds and Encoding}
\begin{figure}[h]
 \centering \includegraphics[width=0.65\linewidth]{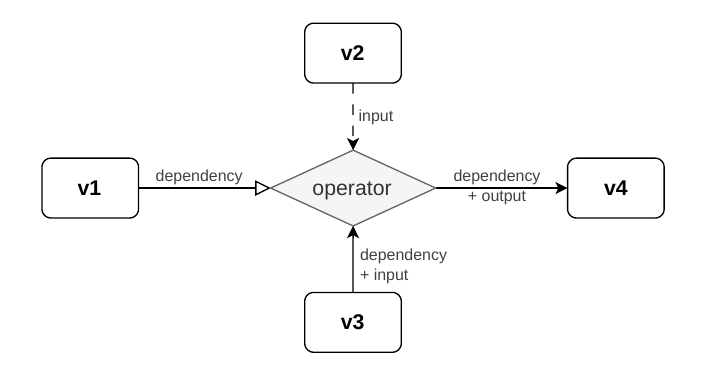}
 \caption{An arrow with an empty head indicates a dependency only; a dotted arrow indicates an input relationship only; an unadorned arrow indicates both a dependency and an input/output relationship.  \label{fig:arrows}}
\end{figure}

A dependency graph in \name consists of two types of nodes: \variable{variable} and \operator{operator}. In this paper's figures, we represent variable nodes as rounded rectangles, and operator nodes as diamonds. An edge always connects a variable and an operator, and no edges exist between two variables or between two operators. 
Figure \ref{fig:arrows} provides a legend for interpreting different types of dependency graph edges:
\begin{itemize}
    \item An arrow with an empty head indicates a dependency: any change in \variable{v1} will invoke the operator, but the operator does not take \variable{v1} as input;
    \item A dotted arrow indicates an input relationship: \variable{v2} is an input of the operator, but a change in \variable{v2} does not invoke the operator;
    \item An unadorned arrow indicates both a dependency and an input/output relationship: \variable{v3} serves as an input to the operator, which outputs \variable{v4}, any change in \variable{v3} will also invoke the operator. 
\end{itemize}
  The motivation for introducing these different types of edges is to reduce redundant computation during change propagation. For instance, there may exist multiple paths connecting an upstream node to a downstream operator, by excluding the input only edges from the traversal, we can avoid invoking the operator multiple times. 

\subsubsection{Bounds}

\begin{figure}[h]
 \centering \includegraphics[width=\linewidth]{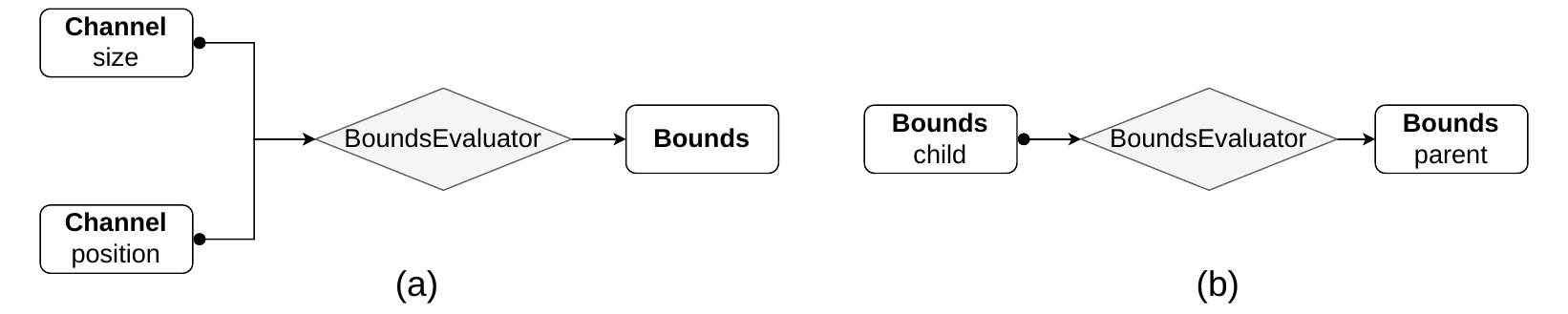}
 \caption{The Bounds Patterns for (a) a mark, and (b) a collection. A circle at the arrow start point indicates one or more input variables.}
    \label{fig:bounds}
\end{figure}

Updating the bounding box of a visual element is essential whenever an interaction modifies its position or size. Figure \ref{fig:bounds} shows two patterns of bounds computation: for a \mk{mark}, a \operator{BoundsEvaluator} takes one or more size \variable{Channel} variables (\eg width and height for rectangle, radius for circle) and one or more position \variable{Channel} variables (\eg x- and y-positions) as input, and outputs the \variable{Bounds}; for a \group{collection}, a \operator{BoundsEvaluator} takes one or more children's \variable{Bounds} as input, and outputs the collection's \variable{Bounds}.

\subsubsection{Data Scope}
\begin{figure}[ht]
 \centering \includegraphics[width=\linewidth]{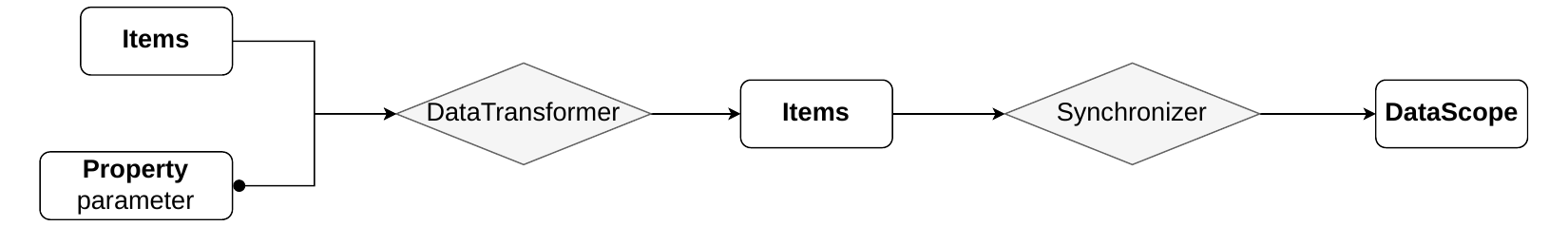}
 \caption{The DataScope Pattern shows that the data scopes of elements depend on (optional) upstream data transformer and synchronizer.}
    \label{fig:datascope}
\end{figure}

The \datascope{data scope} of a visual element (\mk{mark} or a \group{collection})  refers to the data items bound to that visual element for determining the channel values in visual encodings. Figure \ref{fig:datascope} shows that the \variable{items} in an input dataset can be modified by an upstream \operator{DataTransformer} (e.g., a binning operator,  a filtering operator, or a data streaming service) with the corresponding data transformation \variable{parameters} (e.g., the target number of bins, the filtering predicates). Whenever the \variable{items} in an input dataset change, the corresponding elements and their \variable{DataScope} need to be updated by a \operator{Synchronizier}. For example, when the input dataset is filtered by a predicate, the number of marks changes to match the filtered dataset, and their datascopes are updated accordingly.     

\subsubsection{Encoding}

\enc{Visual encoding} is a fundamental component in data visualization, mapping a data attribute to a chosen visual channel of a set of elements. Examples include encoding a quantitative attribute using height in a bar chart, encoding a quantitative attribute using x-position in a lollipop chart, and encoding a nominal attribute using color in a bubble chart. 

Figure \ref{fig:teaser}b shows the design pattern for constructing a dependency graph for a simple visual encoding. Three operators are involved: an \operator{AttributeExtractor}, a \operator{ScaleBuilder}, and an \operator{Encoder}. The \operator{AttributeExtractor} computes the \variable{AttributeValue} for each visual element, based on the input \variable{Attribute}, the \variable{DataScope} of each element, a \variable{Property} specifying how the values should be aggregated if the data scope contains multiple items, and an optional \variable{Transformation} applied to the attribute (e.g., binning a quantitative attribute). The \operator{ScaleBuilder} produces a \variable{Scale} based on the computed \variable{AttributeValue} for each element, one or more scale domain \variable{Property} variables (e.g., whether the domain should start from 0 if the attribute is quantitative), and one or more scale range \variable{Property} variables (e.g., the range extent). The \operator{Encoder} computes and assigns the \variable{Channel} value to each element based on its \variable{AttributeValue} and the \variable{Scale}. The \variable{AttributeValue} is also needed as input to the \operator{Encoder}, and since the \operator{Encoder} is already included in the downstream propagation path, we only create an input relationship (denoted by a dotted arrow) to avoid redundancy in propagation computation.

\begin{figure}[b]
 \centering 
 \includegraphics[width=\linewidth]{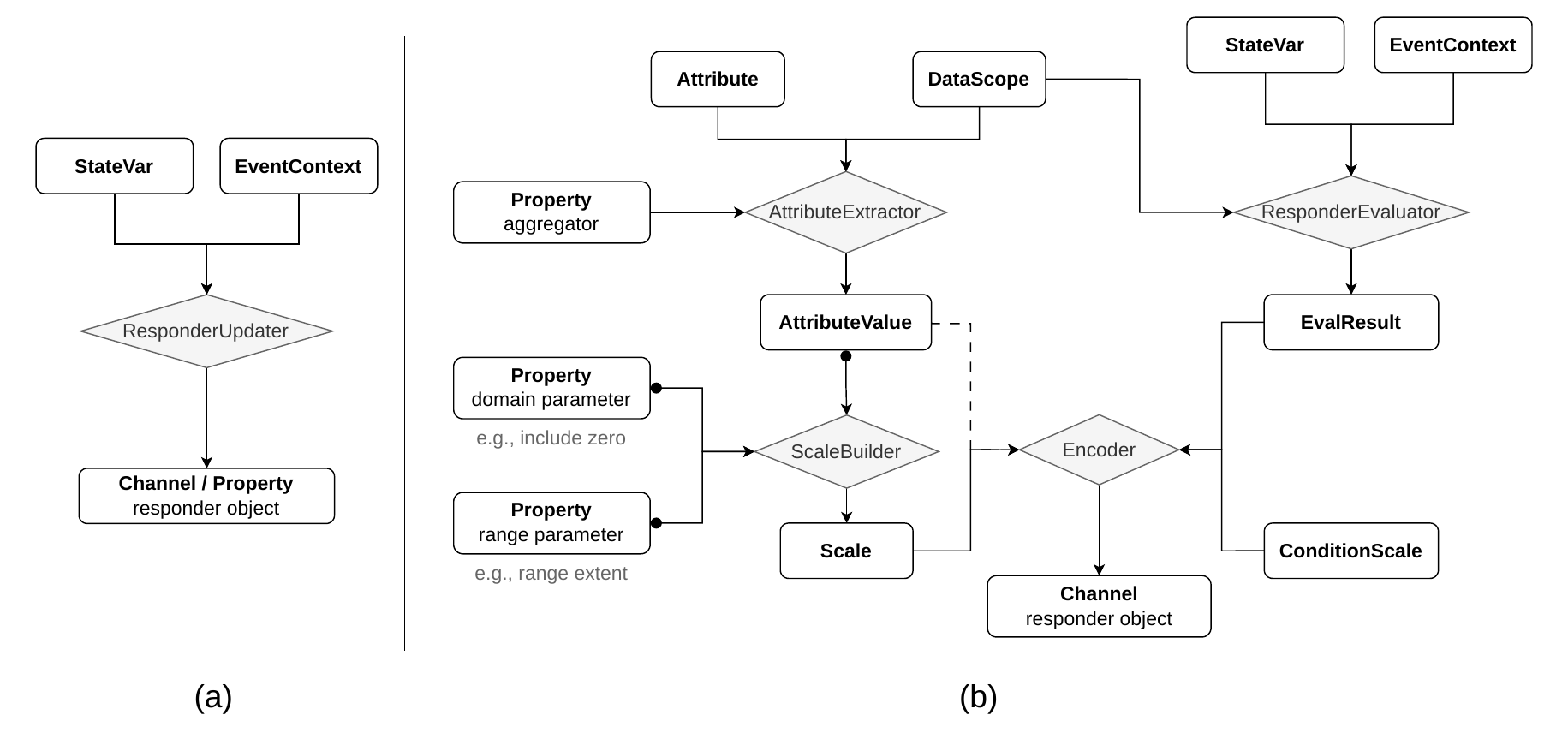}
 \caption{(a) Directly updating a responder based on information in event context. The responder property may belong either to a visualization component or the state context. (b) The Conditional Encoding Pattern where the responder elements' channel values are jointly determined by visual encoding and condition evaluation.  \label{fig:interaction-patterns}}
\end{figure}

\begin{figure*}[ht]
 \centering \includegraphics[width=\textwidth]{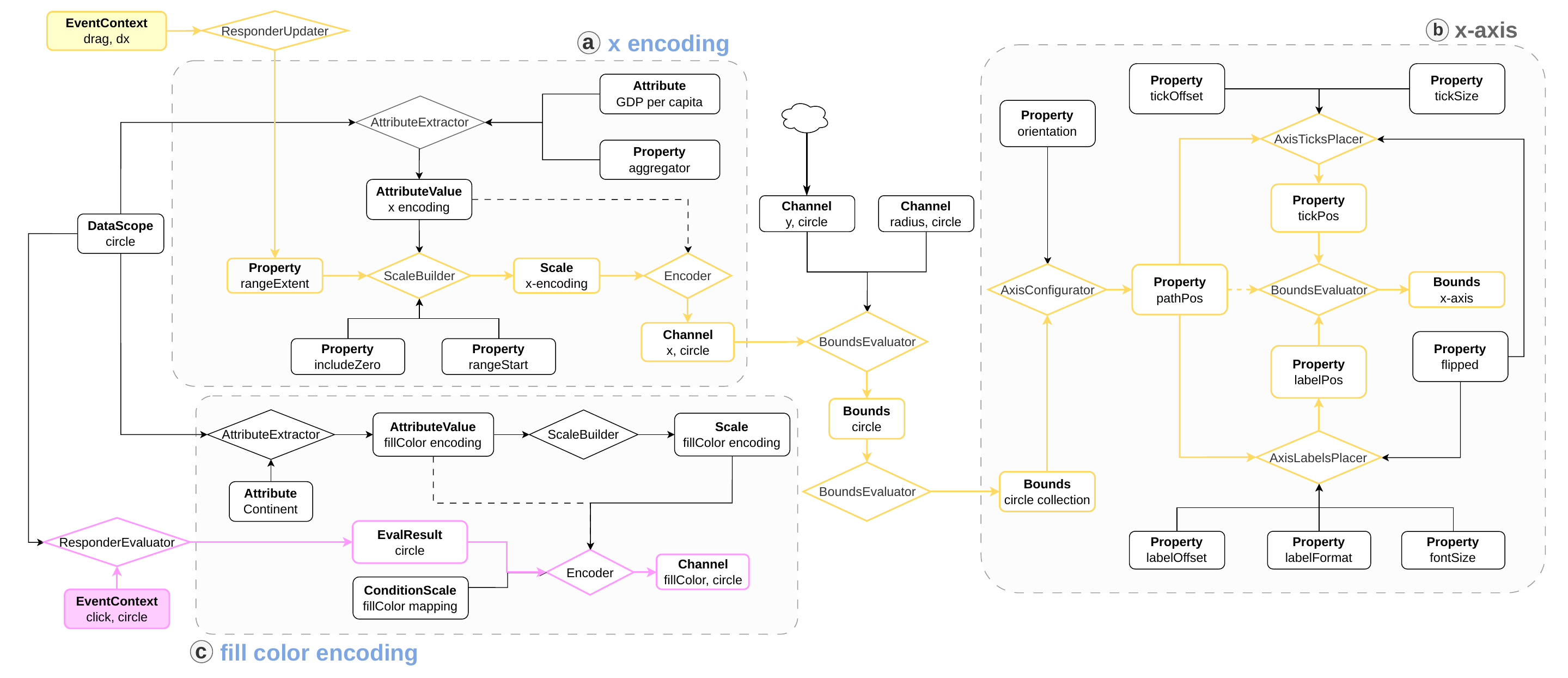}
 \caption{A dependency graph for two interaction designs (E1 and E2) in a scatterplot (Section \ref{sec:interactive_components}). Due to limited space, the subgraphs for y-encoding (represented as the cloud icon), y-axis, gridlines, and legend are omitted.  \label{fig:scatterplot-dg}}
\end{figure*}

\subsection{Dependency Graph Patterns: Direct Property Modification and Conditional Encoding}

\bpstart{Direct Property Modification} Figure \ref{fig:interaction-patterns}(a) shows the dependency graph pattern for interactions where a trigger results in an update of responder property: the information captured in the \variable{EventContext} or a \variable{StateVar} is used by a \operator{ResponderUpdater} to change the \variable{property} or \variable{channel} of the responder. 

\bpstart{Conditional Encoding}
A conditional encoding is often used in conjunction with a regular \enc{visual encoding}: for instance, in E2, the fill color of each mark reverts back to encoding \attr{Continent} when the selection is cleared.  
Figure \ref{fig:interaction-patterns}b shows the pattern for conditional encoding, which includes the visual encoding pattern from Figure \ref{fig:teaser}b. 
A \operator{ResponderEvaluator} operator takes the \variable{EventContext}/\variable{StateVar} and the responder's \variable{DataScope} as input.  
The output is an \variable{EvalResult}, where the value is True, False, or undefined (e.g., when no element is hovered and captured in the \variable{EventContext}). In addition, we also need to specify a \variable{ConditionScale} that maps the \variable{EvalResult} to \variable{Channel} values. 
The \variable{EvalResult} and \variable{ConditionScale} in this conditional encoding play analogous roles to \variable{AttributeValue} and \variable{Scale} in the fill color \enc{visual encoding}. Finally, the \operator{Encoder} operator takes all four inputs and determines the final fill color. When the circle in the \variable{EventContext} is defined, the \operator{Encoder} gives priority to the conditional encoding, setting the \variable{Channel} value based on the \variable{ConditionScale}; otherwise, it defaults to the regular encoding.

\subsection{Dependency Graph: the Scatterplot Examples}
We now show how these component-specific patterns are instantiated and connected to form a complete dependency graph
for the scatterplot examples E1 and E2 introduced in Section \ref{sec:interactive_components}. For clarity, we illustrate only three representative components: 
the \enc{x encoding}, \refmark{x-axis}, and \enc{fill color encoding} (Figure \ref{fig:scatterplot-dg}).

The {\enc{x encoding}} component involves three operators (Figure \ref{fig:scatterplot-dg}a): an \operator{AttributeExtractor} (computing the \attr{GDP~per~capita} \variable{AttributeValue} for each circle based on the circle's \variable{DataScope}, the attribute value will be mapped to x-position), a \operator{ScaleBuilder} (creating a \variable{Scale} and setting its domain and range), and an \operator{Encoder} (using the \variable{Scale} to map each \variable{AttributeValue} to a x-position \variable{Channel} value). For each operator node, the incoming edges connect to the essential variables serving as inputs, and the outgoing edges connect to the output variables. For example, the \operator{ScaleBuilder} operator requires the \variable{AttributeValue} for each circle and additional \variable{Property} variables (e.g., whether the domain starts from zero, the starting x position, and range extent).

In a similar vein, for the \refmark{x-axis}, a different set of operators is involved: \operator{AxisConfigurator}, \operator{AxisTicksPlacer}, and \operator{AxisLabelsPlacer} (Figure \ref{fig:scatterplot-dg}b). Each operator has input and output variables. For instance, the \operator{AxisConfigurator} computes the path position \variable{Property} of the x-axis, which depends on the \variable{Bounds} of the \group{collection} of \mk{circles} and the orientation \variable{Property} of the axis (\ie top or bottom). 

The \enc{x encoding} subgraph and the \refmark{x-axis} subgraph are connected through two \operator{BoundsEvaluator} operators: the first one computes the \variable{Bounds} for each \mk{circle} based on its position (x and y coordinates) and radius; the second one computes the \variable{Bounds} of the \group{collection} based on the bounding boxes of individual \mk{circles}. 

For the \enc{fill color encoding}, we have the same types of variables and operators for the  component as those used for the \enc{x encoding}. In addition, we specify the variables and operators for a \textit{conditional encoding} used in E2 (Figure \ref{fig:scatterplot-dg}c): the interaction is triggered by a click event, with the clicked mark captured by the \variable{EventContext} variable (highlighted in purple). This information serves as input to a \variable{ResponderEvaluator} operator, which compares each responder (\ie the circles) to the clicked circle. The evaluation result \variable{EvalResult} serves as an input to the \operator{Encoder}. In addition, we need to specify a \variable{ConditionalScale} that maps evaluation results to fill colors. The final fill color is jointly determined by the original fill color encoding and the \textit{conditional encoding}.

\subsection{Interactive Behavior as Change Propagation}
\label{sec:interaction-propagation}

The dependency graph provides a systematic way to compute the initial state of the static visualization: starting from the \variable{DataScope} variable, we can invoke all the downstream operators to compute the channel values and axis positions. More importantly, interactive behaviors can be realized using the same dependency graph setup. 

The interactive behavior in E1 is triggered by a drag event on the x-axis, and the specific information about the event (e.g., dx, dy)  is captured by an \variable{EventContext} node (highlighted in yellow). Starting from this node, we propagate the initial change along the directed edges of the dependency graph, invoking each operator to re-evaluate output variables. Figure \ref{fig:scatterplot-dg}  highlights the relevant operators and variables in yellow. The configuration of the dependency graph ensures that only downstream variables that depend on the changed variable are re-evaluated; unrelated portions of the graph (e.g., the \variable{AttributeValues}) remain unchanged. 

Stateful interactions such as E3 follow the same propagation mechanism. Updating a state variable triggers downstream operators that depend on that variable, allowing persistent interaction state to influence subsequent conditional encodings without introducing additional execution mechanisms.  

For E2, whenever a click event occurs, the operators along the downstream propagation path from the \variable{EventContext} node (highlighted in purple) are invoked, dynamically updating each circle's fill color.  

\subsection{More Dependency Graph Patterns}
\label{sec:more_patterns}
In this section, we go beyond the scatterplot examples and present additional dependency graph patterns for common components in data visualization. Together, these patterns support a wide range of visualizations and interactions. 

\subsubsection{Sharing a Scale between Encodings}
\label{sec:sharing_scale}
\begin{figure}[ht]
  \centering  \includegraphics[width=\linewidth]{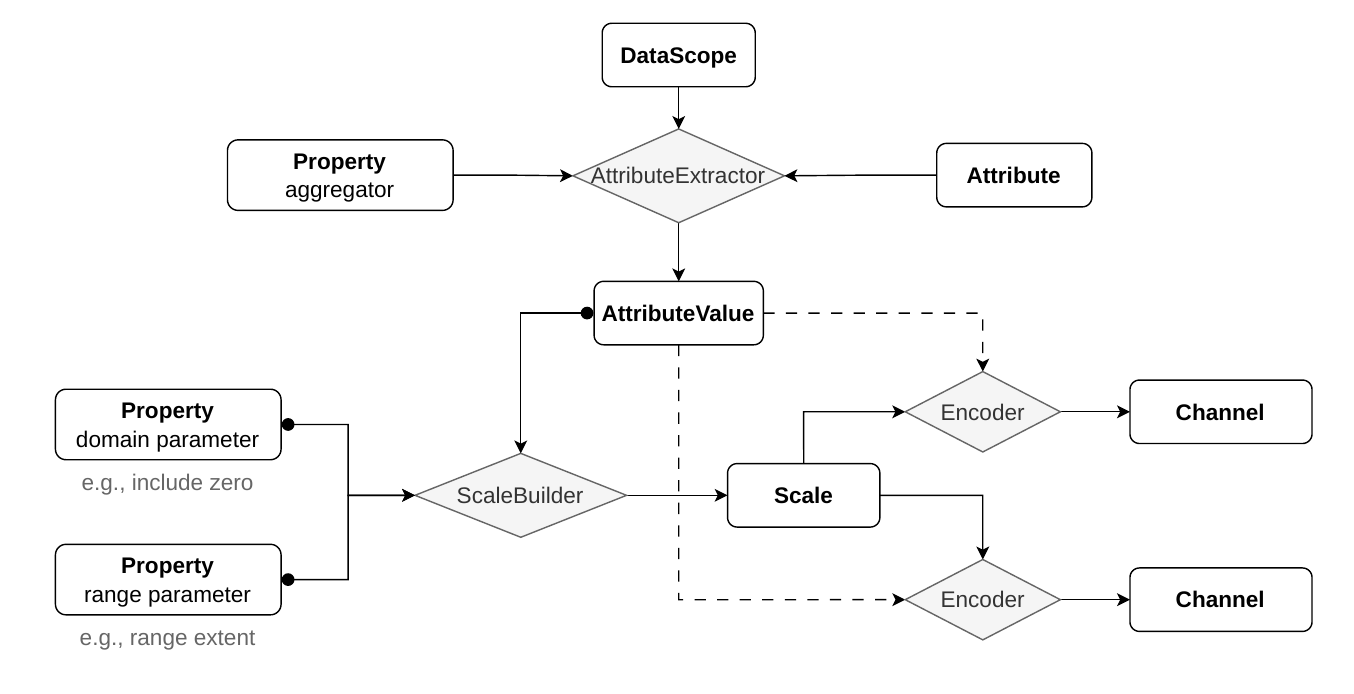}
    \caption{The Shared Scale Pattern. Compared to \Cref{fig:teaser}b, multiple Encoder operators are created, one for each visual encoding. These operators share the same input Scale variable.}
    \label{fig:share-scale}
\end{figure}

In certain data visualization designs, multiple encodings can share the same scale.
Examples include box plots, bullet charts, dumbbell charts, Gantt charts, range charts, and waterfall charts.
For instance, in a waterfall chart showing the change in sales between the current and previous months, there are two \enc{visual encodings}: the y-position of the top bar segment encodes the previous month's sales value, and the y-position of the bottom bar segment encodes the current month's sales value. These two encodings need to share the same scale to ensure that the sales values are encoded consistently. We can extend the dependency graph in Figure \ref{fig:teaser}b to the one in Figure \ref{fig:share-scale} to support this design, where the \variable{Scale} created by the \operator{ScaleBuilder} serves as an input variable for multiple \operator{Encoder} operators. Whenever an upstream variable is changed (e.g., the range extent, the data attribute), the change will be propagated to invoke the respective operators to update the downstream variables.

\subsubsection{Layout}
\begin{figure}[ht]
  \centering \includegraphics[width=0.7\linewidth]{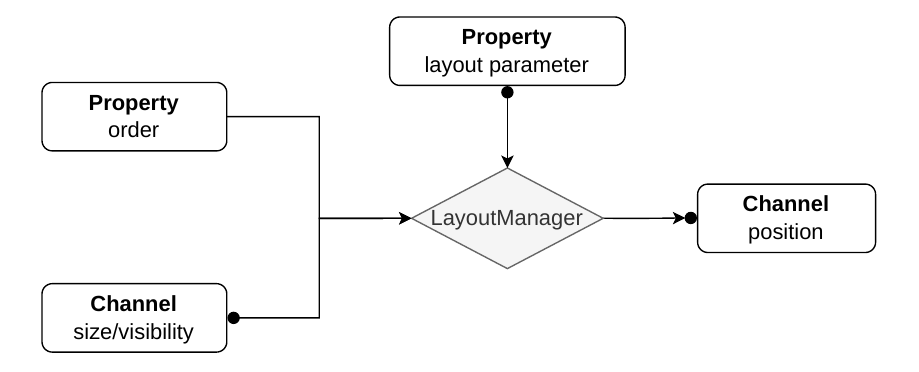}
    \caption{The Algorithmic Layout Pattern.
    A circle at the arrow start or end point indicates one or more variables.}
    \label{fig:layout}
\end{figure}

Besides \enc{position encodings}, the spatial arrangement of visual elements is often achieved through \layout{layouts}. 
Example charts include waffle chart, stacked area chart, and circle packing chart.
Figure \ref{fig:layout} shows the dependency graph pattern for such layouts. Using a waffle chart as an example, the position \variable{Channel} variables (x position and y position) of each rectangle mark are computed by a \operator{LayoutManager} operator of the type ``Grid''. 
The inputs to this operator include one or more size (i.e., the width and height of each mark) and visibility \variable{Channel} variables, one or more \variable{Property} variables representing the layout parameters (e.g., direction of the grid layout, the row/column gaps), and the order \variable{Property} specifying the order of rectangle marks in the collection. Depending on the type of layout used in a chart, the input variables will vary. Whenever the value of an input variable is changed, the operator will be invoked to update the position of each mark. 

\subsubsection{Constraint}
In addition to \enc{position encodings} and \layout{layouts}, \constr{constraints} can also determine the spatial arrangement of visual elements. For example, a diverging bar chart, we can define an \constr{alignment} constraint to align stacked bars by a neutral value, and an \constr{affixation} constraint where the text labels are affixed to the corresponding marks at the center.

\begin{figure}[ht]
  \centering
\includegraphics[width=\linewidth]{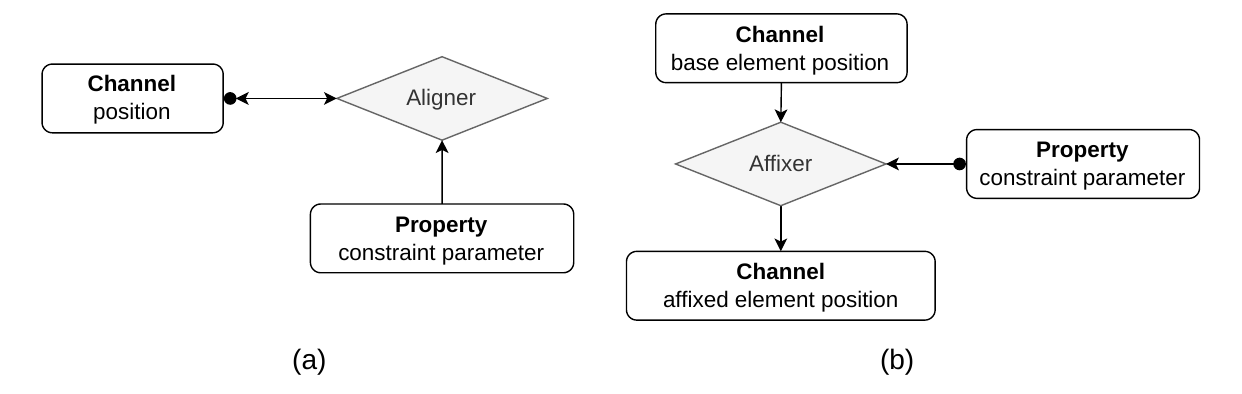}
    \caption{The Alignment Pattern and the Affixation Pattern.}
    \label{fig:constraint}
\end{figure}

Figure \ref{fig:constraint}a shows the pattern for the \constr{alignment} constraint. Unlike the other one-way operators we have seen in this section, an \operator{Aligner} operator is multi-way: one or more position \variable{Channel} variables connected to it can serve as either input or output. In the diverging bar chart example, we have \variable{Channel} variables representing the positions of bars with the neutral value. Any change in one of these variables will invoke the \operator{Aligner} operator and update the other variables. In addition, a \variable{Property} variable representing the anchor parameter serves as the input to the operator. Any change in the anchor property (\eg align middle instead of right) will invoke the \operator{Aligner} operator too.
Figure \ref{fig:constraint}b shows the pattern for the \constr{affixation} constraint, which is defined as a one-way operator in \name. The inputs are one or more affixation parameters (e.g., base element anchor, affixed element anchor, offset) and the base element's position \variable{Channel} (\eg x-position of a bar), and the output is the affixed element's position \variable{Channel}. 

\subsubsection{Nested Collections}
Some visualizations contain nested structures, where multiple \group{collections} of \mk{marks} are children of a higher-level \group{collection}. Examples include but are not limited to grouped/stacked bar charts, mosaic plots, rose charts, and small multiples of simple charts. The spatial arrangement of elements in the lower and higher-level collections can be done either through \enc{position encodings} or \layout{layouts}. 
Appendix B presents the dependency graph patterns for two kinds of nested collections: nested layouts, and encoding-in-layout.  

%% file: tex/4.mascot.tex
\section{Implementation: Mascot.js}
\label{sec:implementation}
We have implemented \name in
the JavaScript library Mascot.js, extending the original library with APIs for specifying interaction components and an execution engine for constructing and maintaining dependency graphs. The demos are available online at 
\href{https://mascot-vis.github.io/gallery/?behavior=interactive}{https://mascot-vis.github.io/gallery/?behavior=interactive},
best viewed in Chrome.

The original Mascot.js \cite{liu_manipulable_2025,liu_atlas_2021} does not support interactivity specification, and focuses on the construction and manipulation of static visualizations. It 
covers all the core static components identified in Table \ref{tbl:components}, and provides APIs to initialize marks, create collections by applying generative operations like repeat and divide, apply layouts, encode data attributes through visual channels, and specify constraints. It also provides APIs to directly access and modify the properties of these components. Lines 1-14 in Listing \ref{code:e1e2} show how to construct the static scatterplot in Figure \ref{fig:scatterplot} using the original Mascot.js.

In the new Mascot.js, we add APIs for specifying trigger, responder, evaluator and updater, and implement the dependency graph model in \name to support reactive behaviors. Programmers specify only the four interaction components. The dependency graph is constructed automatically by Mascot.js from the interaction specification together with the semantic visualization components already present in the scene. This separation allows programmers to reason entirely in terms of interaction semantics, while dependency-graph construction and maintenance remain transparent.

\subsection{APIs for Specifying Interaction Components}
To support interaction specification based on the \name abstraction, we implemented a new function:

\vspace{1mm}
\noindent\colorbox{white}{\inlineCode{msc.activate(trigger, responder, evaluatorFn, updaterFn)}}
\vspace{1mm}

Lines 16-36 in Listing \ref{code:e1e2} show the code to specify interactions in the scatterplot example from Sec. \ref{sec:interactive_components}: E1 (dragging x-axis to change range extent, lines 16-21), and E2 (click on a circle to highlight circles with the same \attr{continent} value, lines 23-36). Due to space limitations, we include the complete Mascot.js implementation of E3 in Appendix C. The example combines multiple interaction behaviors, including direct property modification and conditional encoding, and demonstrates how persistent interaction state is specified using the state context.

\begin{lstlisting}[style=js,caption={Specifying the interactive scatterplot in Sec. \ref{sec:abstraction}: programmers only need to specify the visualization and interaction components, they do not need to understand and construct the dependency graphs.},label={code:e1e2},numbers=left]
let scn = msc.scene();
let dt = await msc.csv("datasets/csv/gdp-lifeExp.csv");
let circle = scn.mark("circle", {radius: 6, x: 100, y: 80});
let collection = msc.repeat(circle, dt, {attribute: "Country"});

let xEnc = msc.encode(circle, {attribute:"GDP/PC", channel:"x"});
msc.encode(circle, {attribute:"life expectancy", channel:"y"});
msc.encode(circle, {attribute:"continent", channel:"fillColor"});

let xAxis = scn.axis("x", "GDP/PC", {labelFormat:".2s"});
scn.axis("y", "life expectancy");
scn.legend("fillColor", "continent", {x: 600, y: 250});
scn.gridlines("x", "GDP/PC");
scn.gridlines("y", "life expectancy");
     
let trigger1 = {event: "dragX", source: xAxis},
     responder1 = {object: xEnc,  properties: ["rangeExtent"]};
let updater1 = (evalResult, evtCtx, stateCtx, respObj) => {
    respObj.rangeExtent += evtCtx.get("dx");
};
msc.activate(trigger1, responder1, undefined, updater1);

let trigger2 = {event: "click", source: circle},
    responder2 = {object: circle, properties: ["fillColor"]};
let evaluator = function(evtCtx, stateCtx, respObj) {
    let source = evtCtx.get("element");
    return !source || 
        source.datum["continent"] === respObj.datum["continent"];
};
let updater2 = (evalResult, evtCtx, stateCtx, respObj) => {
    if (!evalResult) {
        respObj.fillColor = '#eee';
    }
};
msc.activate(trigger2, responder2, evaluator, updater2);

msc.renderer("svg", "svgElement").render(scn);
\end{lstlisting}

The first two parameters of the \inlineCode{activate} method are always required: the \inlineCode{trigger} specifies the \textit{event} and \textit{source}; and the \inlineCode{responder} specifies the \textit{object} and \textit{properties} (multiple properties may be specified). 
Lines 15-16 define the trigger and responder for E1, and lines 22-23 do the same for E2. The third parameter \inlineCode{evaluatorFn} is only required when specifying conditional encoding. It is a function that accepts three parameters: event context (\inlineCode{evtContext}), state context (\inlineCode{stateContext}), and responder object (\inlineCode{respObj}), and returns a boolean result. This function is used to create a  \operator{ResponderEvaluator} operator in the dependency graph. Lines 24-28 specify an \inlineCode{evaluatorFn} for E2 that evaluates each circle based on its \attr{Continent}
value, where the value of \inlineCode{respObj} is defined in the \inlineCode{responder2}. The fourth parameter \inlineCode{updaterFn} is always required for specifying how to update a responder's properties either directly (corresponding to \operator{ResponderUpdater}) or based on the evaluation results in a conditional encoding (corresponds to \variable{ConditionScale}). It is a function that accepts four parameters: evaluator result (\inlineCode{evalResult} - the boolean result returned by the evaluator function, if applicable), event context (\inlineCode{evtContext}), state context (\inlineCode{stateContext}), and responder object (\inlineCode{respObj}). Lines 17-19 show the specification of how the dragging event directly affects the range extent of the \enc{x encoding} in E1, and lines 29-34 show how to update the fill color and opacity of the circles in E2. 

When a visualization involves multiple interaction behaviors, the interaction components can be reused. For instance, the same \inlineCode{trigger} may apply to more than one \inlineCode{responder}. 

\subsection{Dependency Graph Construction}
The set of patterns presented in Sec. \ref{sec:patterns} serves as a blueprint to be closely followed for systematic and consistent constructions of dependency graphs. When formulating the patterns, we first identify the input/output relationships between variables and operators, and these relationships are represented as input/output edges (dotted arrow in Figure \ref{fig:arrows}). We then decide if each of these edges can be promoted to be a dependency+input/output edge (unadorned arrow in Figure \ref{fig:arrows}) by examining the overall topology: given an input/output edge that directly connects two nodes, if there does not exist a path that connects two nodes through additional intermediate nodes, the edge will be promoted. Finally, we add dependency-only edges (empty head arrow in Figure \ref{fig:arrows}) to join disconnected subgraphs. Such edges are useful in nested collections (Appendix B). 

Following these patterns, Mascot.js incrementally and automatically constructs the dependency graph for a scene, driven by the procedural grammatical specifications. A dependency graph $dg$ is instantiated along with the initialization of a scene (line 1, Listing \ref{code:e1e2}). Whenever an operation is applied (e.g., mark instantiation, repeat, encode), the dependency graph is updated with new nodes and edges. For example,  instantiating a new \mk{circle mark} (line 3) will create a subgraph based on the Bounds Pattern in Figure \ref{fig:bounds}(a); repeating the \mk{circle} mark will create a subgraph based on the Bounds Pattern in Figure \ref{fig:bounds}(b).    

To specify an interactive behavior, programmers only need to specify the trigger, responder, evaluator, and updater.  No knowledge of the underlying dependency graph is needed. Mascot.js automatically constructs the required dependency graph based on the Interaction Patterns in Figure \ref{fig:interaction-patterns} using 
the four specified interaction components. Algorithm \ref{algo:activate} describes how the dependency graph is updated when \inlineCode{msc.activate} is called. Since the dependency graph construction is incremental, the encoding subgraph is usually already created as part of the conditional encoding graph to be constructed. For example, line 8 in Listing \ref{code:e1e2} specifies a \enc{fill color encoding}, which results in an encoding subgraph constructed based on the Encoding Pattern in Listing \ref{fig:teaser}(b). When the conditional encoding interaction is specified in line 36, The algorithm will find relevant \variable{variables} and \operator{operators} in the graph, and create new nodes and edges as needed.

\FloatBarrier
\begin{algorithm}[!t]
\caption{\texttt{msc.activate}\\
A variable ends with \textit{Var} (e.g., \textit{ctxVar}); \\
An operator ends with \textit{Op} (e.g., \textit{evalOp}); \\
\textit{dg.getVariable}: finds an existing variable node;\\
\textit{dg.findInOperator}: finds an existing operator node;\\
\textit{dg.createOperator}: creates a new operator node;\\
\textit{dg.connect}: creates an edge connecting a variable and an operator. By default, the edge is a dependency + input/output edge;
}
\KwIn{\textit{trigger},\ \textit{responder},\ \textit{evaluator},\ \textit{updater}  }
\KwOut{updated dependency graph}
\textit{dg} $\leftarrow$ \textit{scene.dependencyGraph}

\If{\textit{trigger.source} is stateVar}{
\textit{ctxVar} $\leftarrow$ \textit{dg.getVariable}(\texttt{\small VarType.State}, \textit{trigger})
} 
\Else {
\textit{ctxVar} $\leftarrow$ \textit{dg.getVariable}(\texttt{\small VarType.EvtCtx}, \textit{trigger})
}

\If{\textit{evaluator} exists}{
    \textit{outVar} $\leftarrow$ \textit{dg.getVariable}(\texttt{\small VarType.EvalOutput}, \textit{evaluator})
    
    \textit{evalOp} $\leftarrow$ \textit{dg.findInOperator}(\texttt{\small OpType.RespEvaluator}, \textit{outVar})
    
    \If{not \textit{evalOp} exists}{
        \textit{evalOp} $\leftarrow$ \textit{dg.createOperator}(\texttt{\small OpType.RespEvaluator}, \textit{evaluator})
    }
    \textit{dg.connect}(\textit{ctxVar, evalOp})
    
    \textit{dg.connect}(\textit{evalOp, outVar})

    \ForEach{\property~$\in$ \textit{responder.properties}}{
        \textit{pVar} $\leftarrow$ \textit{dg.getVariable}(\texttt{\small VarType.Property}, \property, \textit{responder})

        \textit{encOp} $\leftarrow$ \textit{pVar.inOperator}
        
        \If{\textit{encOp} exists}{
            \textit{dg.connect}(\textit{outVar}, \textit{encOp})
        }
        \Else {
            \textit{encOp} $\leftarrow$ \textit{dg.createOperator}(\texttt{\small OpType.Encoder})

            \textit{dg.connect}(\textit{outVar}, \textit{encOp})

            \textit{dg.connect}(\textit{encOp}, \textit{pVar})
        }
        \textit{condScaleVar} $\leftarrow$ \textit{dg.getVariable}(\texttt{\small VarType.ConditionScale}, \textit{updater}, \textit{responder})

        \textit{dg.connect}(\textit{condScaleVar}, \textit{encOp})
    }
} 
\Else{
    \textit{updateOp} $\leftarrow$ \textit{dg.createOperator}(\texttt{\small OpType.RespUpdater}, \textit{updater}, \textit{responder})

    \ForEach{\property~$\in$ \textit{responder.properties}}{
        \textit{pVar} $\leftarrow$ \textit{dg.getVariable}(\texttt{\small VarType.Property}, \property, \textit{responder})

        \textit{dg.connect}(\textit{ctxVar}, \textit{updateOp})

        \textit{dg.connect}(\textit{updateOp}, \textit{pVar})
    }
}
\label{algo:activate}
\end{algorithm}

\subsection{Performance Optimization}
We optimize the performance of change propagation along dependency graph paths using the following strategies. First, only dependency edges are included in the path traversal, and input only edges are excluded. Second, an upstream change can propagate to multiple downstream variables that serve as inputs to the same operator. For instance, in Figure \ref{fig:scatterplot-dg}, a change in \variable{EventContext} is propagated to \variable{tickPos} and \variable{labelPos} (paths highlighted in yellow), both serving inputs to the \operator{BoundsEvaluator}, causing this operator to be invoked twice. To avoid such redundant computations, we do a breadth-first traversal to add all the operator invocations to a list, remove duplicated operators from that list, and then invoke the remaining operators in the list sequentially. 

The dependency graph is also useful for improving rendering performance. When propagating an initial variable change along the graph paths, we assign a ``dirty flag'' to each visual element (e.g., mark, axis) whose visual properties are updated as a result of operator invocation. After the end of each propagation, the scene needs to be re-rendered. We can selectively update the visual elements with the dirty flags only. After each round of re-rendering, the dirty flags are reset.

%% file: tex/5.related_work.tex
\section{Related Work} 
\subsection{Abstractions of Visualization Interactivity}
Existing approaches to authoring interaction in data visualization assume a range of input static charts and abstractions of interactivity. The SVG format is a popular choice for representing static charts, and interaction is accomplished through event handlers on selected SVG elements in tools such as ProtoVis \cite{bostock_protovis_2009}, D3 \cite{bostock_d3_2011}, and VisDock \cite{choi_visdock_2015}. Liu \etal \cite{liu_spatial_2024} propose a spatial model for manipulating static visualizations, and the focus is on interactions that can be simulated using physics-based constraints. While SVG provides a flexible representation of visual elements, it lacks the high-level semantic abstractions needed to conveniently specify diverse interactive behaviors, requiring programmers to coordinate responses through low-level event handlers.

Visualization grammars provide higher-level abstractions for specifying visualizations and their behaviors. Vega \cite{satyanarayan_reactive_2016} defines interaction primitives such as event streams, signals, and transforms, and  Vega-Lite \cite{satyanarayan_vega-lite_2016} provides higher-level abstractions for specifying selections, predicates, and transforms. 
In this work, we use the term grammar to refer to a structured set of abstractions together with the information-flow rules  governing how they interact. \name adopts this perspective by defining four interaction components, two forms of context, and the information-flow rules that govern their interactions. Libra \cite{zhao_libra_2025} provides an interaction model for visualizations created using different libraries. Unlike Libra which represents interactions through layers, instruments, interactors, and services, Interactive Mascot organizes interactions around semantic scene components and their information-flow relationships. 
Among these different abstractions, \name is closest to Vega-Lite. Both provide high-level interaction grammars that separate interaction semantics from low-level execution mechanisms. 
Unlike Vega-Lite which organizes interaction around data-centric abstractions such as   selections and transformations, \name's scene-centric approach allows scene elements to participate directly in interaction specification and execution rather than being encapsulated behind data-centric abstractions.

\subsection{Modeling Interactivity Using Graphs}
Early work in graphical user interface toolkits such as Garnet and Amulet \cite{vander_zanden_lessons_2001,myers_amulet_1997,myers_garnet_1990} has used dataflow graphs to keep track of dependencies among variables and constraints for the development of UI interactive behaviors. The Stencil model \cite{cottam_stencil_2008} treats user interaction as a stream of event descriptors, which is used together with input data to control the properties of visual displays. In Reactive Vega \cite{satyanarayan_reactive_2016}, the Vega runtime parses a high-level declarative specification and constructs a dataflow graph that captures the dependencies between operators, which perform computations such as input data processing, interaction handling, or scene graph construction. Two types of edges exist in the graph: parameter edges that indicate value dependency, and pulse edges that indicate flow of data objects. In terms of topology, the dependency graphs in \name are more similar to those used in Garnet and Amulet than those in Vega. In Vega dataflow graphs, the nodes are all dataflow operators, whereas the nodes in \name dependency graphs are either variables or operators, and the variable nodes represent component properties at a finer level of granularity. This variable-operator pattern has been used in information visualization frameworks \cite{heer_software_2006}. Such a pattern makes visualization processing reconfigurable, where updating a component only involves the necessary building blocks. The novel contribution of \name is a set of dependency graph patterns specifying the input, output variables and operators, as well as different types of edges connecting the variables with the operators. These patterns serve as reusable graph templates associated with semantic visualization components. Once the scene components are identified, these templates can be systematically instantiated to produce an executable dependency graph.

%% file: tex/6.evaluation.tex
\section{Evaluation of \name}
We evaluate \name and its implementation in terms of performance and usability. 
The performance evaluation measures the interactive frame rates. We conduct a qualitative study with Vega/Vega-Lite users to understand the strengths and limitations of Mascot.js as an interaction authoring tool.




\begin{figure*}[hb]
 \centering \includegraphics[width=\linewidth]{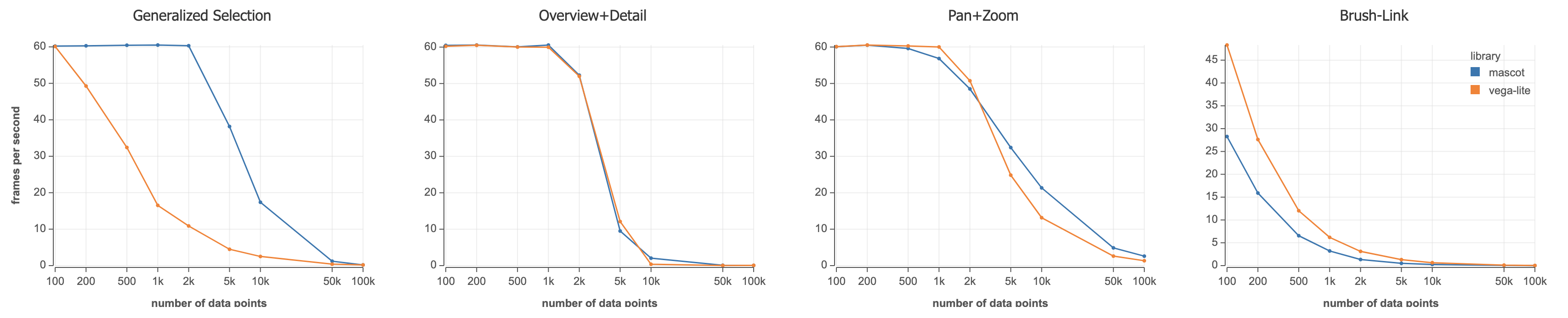}
 \caption{Performance benchmark results across four interaction designs and nine different dataset sizes. The x-axis is in log scale.  \label{fig:benchmark}}
\end{figure*}

\subsection{Performance Benchmark}
\label{sec:benchmark}
To evaluate run-time performance, we compare to Vega-Lite (v6.1.2).
We compare the frame rate in \textit{four} interaction designs implemented using Mascot.js and Vega-Lite with \textit{nine} different dataset sizes (100, 200, 500, 1K, 2K, 5K, 10K, 50K, 100K). The four interaction designs are: scatterplot selection by attribute value (E2), time series  overview + detail (E4), scatterplot matrix (SPLOM) brushing and linking (E5), and scatterplot pan and zoom (E6). 
The last three interactions were also used in benchmark tests in previous work \cite{satyanarayan_reactive_2016,zhao_libra_2025}.

The visualizations implemented using both libraries are all rendered in SVG format. We use Puppeteer \cite{the_chrome_browser_automation_team_puppeteer_2025} to programmatically simulate mouse events. For each combination of library, dataset size, and interaction design, the mouse event simulations are run across 100 iterations. The exact simulation logic depends on the interaction design (details are included in the README file in the supplemental materials). We record the time taken to update a frame for each of the iterations, and derive the corresponding frame rate. We ran the benchmarks using Puppeteer version 24.14.0, Chromium version 138.0.7204.157 on a 2.6 GHz 6-Core Intel Core i7 MacBook Pro (OS X 15.7.3) with per-core 256K L2 caches, 12MB L3 cache, 16 GB 2667 MHz DDR4 memory and a PCI Express AMD Radeon Pro 5300M 4 GB graphics card.

Overall, Mascot.js  maintains interactive performance comparable to Vega-Lite (Figure~\ref{fig:benchmark}). Mascot and Vega-Lite have comparable frame rates in the overview + detail (E4) and pan + zoom (E6) interactions. Mascot performs better in generalized selection (E2), while Vega-Lite performs better in SPLOM brushing and linking (E5). 

The differing performance characteristics likely reflect the distinct interaction abstractions and execution strategies adopted by the two grammars. Interactions such as generalized selection naturally align with Mascot's scene-centric abstraction, which directly reasons over scene objects rather than data selections. Conversely, brushing and linking naturally aligns with Vega-Lite's data-centric selection model, whereas Mascot evaluates the interaction for every scene objects. A detailed profiling study would be required to precisely attribute the observed differences and is left for future work.

The current implementation of Mascot.js allows evaluators and updaters to be specified as arbitrary JavaScript functions, enabling rich interactions such as custom tooltips and comparison against state variables. A more declarative representation of evaluators and updaters (e.g., \inlineCode{updater = \{false : \{fillColor:"\#eee", opacity:0.5\}\}}) could enable additional optimizations, such as bulk evaluation and updates  of scene objects. The decision to represent evaluators and updaters as functions prioritizes expressiveness and programmability over maximum runtime efficiency. Furthermore, large-scale brushing and linking is often constrained by perceptual scalability rather than interaction latency \cite{liu_immens_2013}.

\subsection{User Study}
\label{sec:user_study}
To better understand the usability and learnability of Mascot.js, we conducted a chart re-production user study, following the protocols from similar studies for evaluating chart authoring tools \cite{chen_mystique_2023,liu_data_2018, shi_piccl_2026}. The study was approved by the university's IRB, and we obtained informed consent from all participants, who signed a consent form (included in the supplemental materials). 

\bpstart{Participants} We recruited 10 participants (6 males, 4 females) through mailing lists and social media. Nine participants had prior experiences using Vega-Lite to create data visualizations: 1-2 years (5), 3-4 years (2), more than 4 years (2). The remaining participant primarily used D3 for interactive visualization development. The participants were computer science graduate students (5), researchers (3), professors (1), and software engineers (1).

\bpstart{Procedure} 
All the study sessions were conducted remotely on Zoom and each lasted about 90 minutes. We first went through two tutorials with each participant: the first tutorial covered the example of dynamically changing the number of columns in a waffle chart based on user input in a spinner widget; the second tutorial covered the example of brushing the x-axis of a scatterplot to highlight marks falling within the brushed range. Through these two examples, we taught the participants the following key concepts: static components in charts, four interaction components in the grammar, and the APIs for specifying static and interactive components. The tutorials were taught in a pre-configured web application along with a Google Slides deck explaining the key concepts (e.g., targets and events supported in Mascot.js) and the corresponding JavaScript code for each of the two examples. The tutorial session lasted about 30-40 minutes.

The participants were then asked to complete three tasks independently: dragging the x-axis area of a Gantt chart to rescale the axis (Task 1); brushing the two y-axes in a slope graph to highlight lines whose endpoints both fall within the brushed ranges (Task 2); changing the data attribute encoded as the marks' x positions in a scatterplot through a dropdown menu (Task 3). These three tasks represented three different intents: reconfigure, select, and encode. Task 1 and 3 did not involve an evaluator but Task 2 required an evaluator definition. The tasks built upon the two tutorials but were more challenging than the tutorials. For each task, participants were given the code for the static chart within the pre-configured web application, a video demonstrating the interaction, a natural language description of the interaction, and a Google Slides deck with structures similar to those used in the tutorials (but without the answers). The slide deck asked the participant to first conceptually identify the four components and then translate the components into code. The participants were asked to think aloud while working on the tasks. Since our focus is to evaluate the \name model abstraction and the interaction features in Mascot.js, we provided relevant Mascot.js APIs for static charts as hints (\eg how to set the range extent of an encoding, how to get the positions of a line's start and end points), and participants could ask questions about the APIs. They could also refer to the tutorials whenever necessary. Coding agents were turned off during the sessions. After finishing the tasks, the participants completed a survey to evaluate various usability and learnability dimensions of Mascot.js on a 7-point Likert scale. We also conducted a short semi-structured interview to get their qualitative feedback on Mascot.js.

The study was conducted using an earlier implementation of Mascot.js that predates the introduction of explicit state contexts. Because the study focused on the learnability and usability of the four core interaction components, and none of the study tasks required stateful interactions, we believe the findings remain representative of the current grammar.

\bpstart{Results} 
The participants successfully completed 29 of 30 tasks. Only one participant failed to complete Task 3 because the session exceeded the planned one-and-a-half hours. Participants were able to correctly \textit{identify the four components} for 27 out of 30 tasks without any assistance, and required help for 2 tasks.  They correctly \textit{translated the four components into JavaScript code} for 21 of the 30 tasks without assistance, and required hints or reminders about API usage for 8 of the tasks. The average completion time for each task is: 13.1 minutes (Task 1), 15.0 minutes (Task 2), 9.8 minutes (Task 3). For the post-study survey, the results are as follows (1 = ``strongly disagree'', 7 = ``strongly agree''): model is easy to understand ($\mu$ = 6.0, $\sigma$ = 0.8), model is easy to remember ($\mu$ = 5.9, $\sigma$ = 1.3), API is easy to learn ($\mu$ = 5.8, $\sigma$ = 0.8), easy to debug ($\mu$ = 4.9, $\sigma$ = 1.3), code has low complexity ($\mu$ = 5.6, $\sigma$ = 0.8), cognitive effort is low ($\mu$ = 5.3, $\sigma$ = 1.3), physical effort is low ($\mu$ = 6, $\sigma$ = 1.2), overall satisfied ($\mu$ = 6, $\sigma$ = 0.7). 
  Appendix D provides detailed information on the survey responses. 

In general, participants found the grammar easy to understand: ``\textit{What I like about the library is it's very intuitive ... here we are, like, manipulating visual elements directly, which makes it more intuitive for me to understand}'' (P9). They also commented positively on the abstraction: ``\textit{All you need to do is just to declare the four components, and then everything else is taken care of ... I really like the level of abstraction here}'' (P1). Several participants noted that while the grammar was initially unfamiliar, it was learnable potentially even for non-experts: ``\textit{Of course, there was a little bit of a learning curve ... but even for people without a lot of technical expertise, they should be able to pick up on this}'' (P4). 

Participants' unfamiliarity with the Mascot API presented a major hurdle in the implementation and debugging process: ``\textit{I'm not sure what are those available properties. Because I'm not familiar with the vocabulary of the new language, so it's not easy to write code}'' (P6). This problem may be addressed with better documentation and more examples: ``\textit{if I am actually working with this library, and if I have a comprehensive documentation where I can look things up.  I think it should be much easier to use}'' (P1).

A few participants found that the API for specifying trigger target could be confusing: when the target was a UI widget, it was referenced by its DOM ID, but when the target was a visualization component, it was referenced by the JavaScript variable. We plan to enforce more consistency in the Mascot.js API for cases like this.

%% file: tex/7.conclusion.tex
\section{Discussion and Future Work}
Visualization programs are cognitive notations and can potentially be compared using quantitative metrics \cite{kruchten_metrics-based_2024}. We decided not to perform a quantitative comparison of code complexity because Mascot.js and Vega-Lite adopt fundamentally different paradigms and syntax. Although both are high-level grammars, Vega-Lite uses a compact JSON-based specification with dedicated interaction primitives, whereas \name is implemented as an embedded JavaScript DSL. Our experience in implementing the interaction examples suggests that Vega-Lite specifications are often more concise for interaction techniques directly supported by its built-in interaction primitives. However, as interactions become more complex, the required logic in Vega-Lite increasingly relies on additional signals, predicates, or a transition to Vega. Interactive Mascot, on the other hand, preserves the same four-component interaction structure for all kinds of interaction techniques. The primary advantage of Interactive Mascot lies not in minimizing code length, but in providing a uniform semantic organization that scales naturally to more complex interaction behaviors. Developing language-independent measures of interaction specification complexity remains an interesting direction for future work.

Visualization grammars can be difficult to debug when they hide low-level execution details from users. This observation has been confirmed in previous work \cite{hoffswell_visual_2016} and in our qualitative study. Future work could investigate debugging and inspection tools for \name. While techniques such as event replay, in-situ annotations and dynamic tables \cite{hoffswell_interactive_2019} are potentially applicable, we believe the abstractions introduced by Interactive Mascot can enable new debugging capabilities such as semantic stepping through interaction components, inspection of event and state contexts, visualization of information flow, and explanations of why interaction effects occur. These opportunities extend beyond debugging reactive execution to debugging the semantics of interaction specifications themselves.

We are also interested in exploring how \name can enable novel interaction authoring interfaces. The concise definitions of trigger and responder may be achieved through direct manipulation interaction such as drag and drop. The formulation of evaluator and updater may be accomplished through visual programming and natural language descriptions.  
The four-component abstraction may also enable reusing and transferring interactive behaviors across visualizations. Analysis and annotations of existing interaction patterns, especially in terms of evaluator and updater functions, can inform efforts that build recommendation systems for interactivity. This requires further research into curating and annotating interactive visualizations that go beyond existing corpora research on static visualizations \cite{chen_visanatomy_2026,ko_natural_2024,chen_composition_2021}.

Animation support is another research direction to pursue. State-of-the-art visualization libraries and tools \cite{bostock_d3_2011,satyanarayan_vega-lite_2016,thompson_data_2021,ge_canis_2020,ge_cast_2021} include support for animation authoring. Recent efforts also try to provide a unified framework for both interaction and animation \cite{zong_animated_2023}. The current implementation of Mascot.js provides basic support for animated transition, where the \inlineCode{msc.activate} function accepts a fifth parameter for animation specifications on the timing and duration of responder update. To unify interaction and animation, the trigger component can potentially be extended to include cases where no user action or event is required. We are especially interested in extending the \name abstraction for different authoring paradigms \cite{thompson_understanding_2020} such as keyframing, procedural animation, and presets/templates.

\section{Conclusion}
We contribute \name, a scene-centric grammar for interactive data visualizations. \name identifies interaction components such as trigger and responder that operate seamlessly with existing abstractions of static visualization components, and defines permissible information flows between these components and the event and state contexts. \name 
employs an execution model based on a set of reusable dependency graph patterns, which realizes interaction as the propagation of changes along dependency paths. \name demonstrates that interaction can be specified around semantic visualization components rather than event-processing and data selection abstractions. We hope this perspective will inform future visualization grammars, authoring systems, and AI-assisted interaction generation.

%% file: tex/appendix.tex
\newpage

\appendix
\section{Event Context Fields}
\label{appendix:evtCtx_fields}
The event context \context~associated with a trigger records a set of relevant information necessary to interpret and process that event. An event context typically captures multiple context fields depending on the type of event. Table \ref{tbl:evtCtx} shows commonly used fields in event contexts for various events. 

\begin{table}[h]  \renewcommand{\arraystretch}{1.28}
    \centering
    \caption{Examples of commonly used events and the corresponding event context fields.}
    \begin{tabular}{p{1.65cm} p{1.6cm} p{3.8cm}}
    \toprule
     \textbf{Context Field} & \textbf{Events} & \textbf{Description} \\
     \midrule
\texttt{x}            & hover, drag, scroll, click & Cursor x position in scene/pixel coordinates \\
\texttt{y}            & hover, drag, scroll, click & Cursor y position in scene/pixel coordinates \\
\texttt{xAttr}        & brush, hover               & Name of the data attribute encoded on the x axis \\
\texttt{yAttr}        & brush, hover               & Name of the data attribute encoded on the y axis \\
\texttt{xCoords}      & brush                      & Brush extent in pixel space $[x_0, x_1]$ \\
\texttt{yCoords}      & brush                      & Brush extent in pixel space $[y_0, y_1]$ \\
\texttt{xVals}        & brush                      & Brush extent mapped to data space $[\min, \max]$ \\
\texttt{yVals}        & brush                      & Brush extent mapped to data space $[\min, \max]$ \\
\texttt{xVal}         & hover                      & Data value of the x attribute at cursor position \\
\texttt{yVal}         & hover                      & Data value of the y attribute at cursor position \\
\texttt{selRows} & brush                      & Data rows whose attribute values fall within the brush \\
\texttt{dx}           & drag                       & x displacement since last drag event \\
\texttt{dy}           & drag                       & y displacement since last drag event \\
\texttt{deltaX}       & scroll                     & Scroll wheel x delta \\
\texttt{deltaY}       & scroll                     & Scroll wheel y delta \\
\texttt{element}      & hover, click               & The hovered or clicked element \\
\texttt{inputVal} & change & the value of input UI widget (e.g., checkbox, text input)\\
     \bottomrule
    \end{tabular}
    \label{tbl:evtCtx}
\end{table}

\section{Dependency Graph Patterns for Nested Collections}
\label{appendix:nested_patterns}

\begin{figure}[H]
 \centering \includegraphics[width=0.95\linewidth]{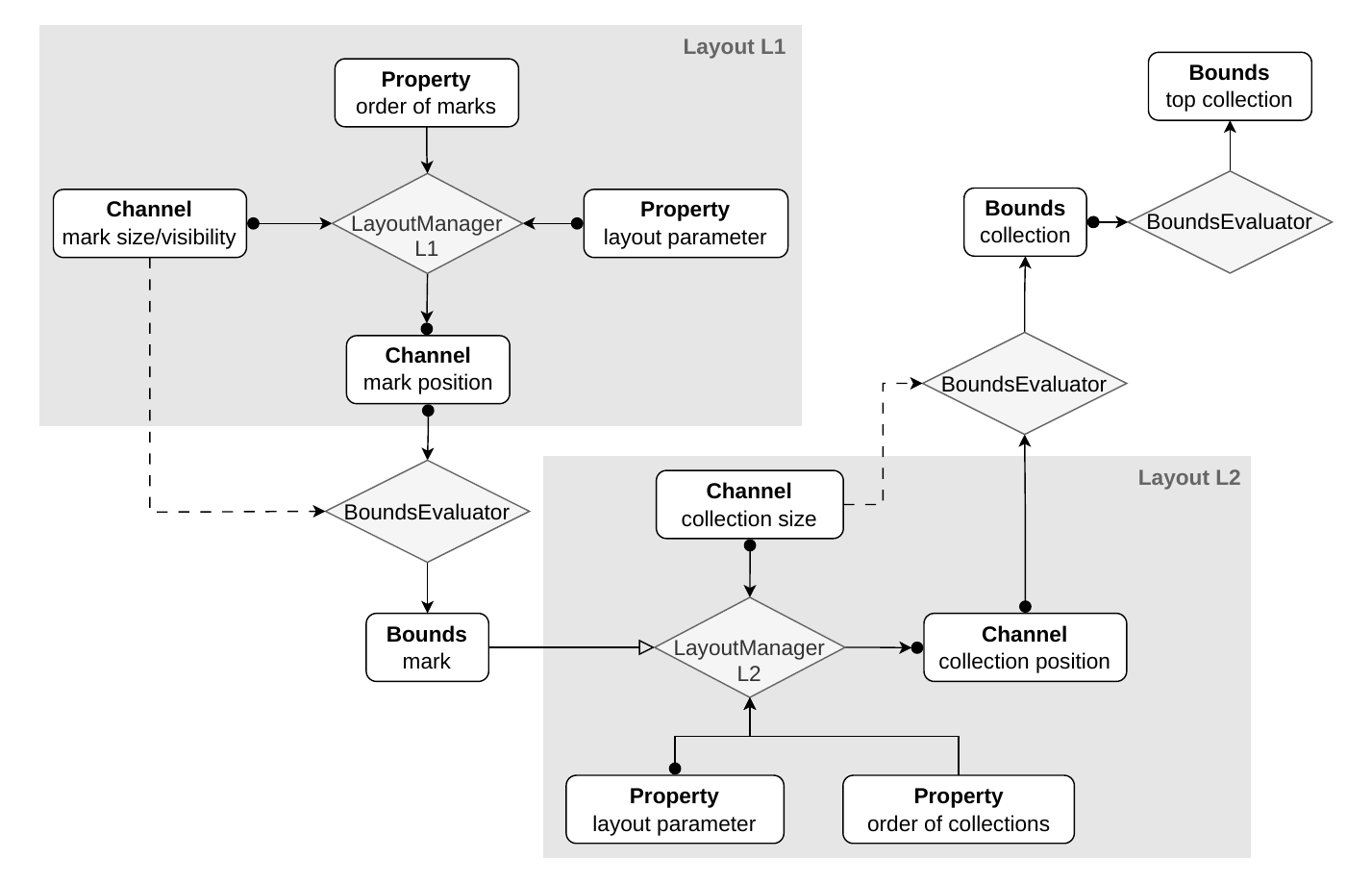}
 \caption{The Nested Layouts Pattern.   \label{fig:nested-layouts}}
\end{figure}

Consider the mosaic plot in Figure \ref{fig:nested-layouts} showing the percentage of male/female employees across five job levels. There are five \group{collections}, each containing two \mk{rectangle marks} color-coded by \attr{gender}. Within each of these \group{collections}, the rectangles are placed in a \layout{vertical stacked layout}. These five \group{collections} are then arranged in a \layout{horizontal stack layout}. Figure \ref{fig:nested-layouts} shows the pattern for such structures with \textbf{nested layouts}. The inputs to the first-level \operator{LayoutManager} L1 operator include the size \variable{Channel} variables (\ie the width and height of each rectangle mark), layout parameters \variable{Property} variables (\eg orientation of stacking), and the order \variable{Property} of the rectangle collection. The output of the operator is one or more position \variable{Channel} variables (\ie the x- and y-positions). This configuration is consistent with the design pattern in Figure \ref{fig:layout}. The second-level \operator{LayoutManager} L2 works similarly, computing the position \variable{Channel} variables for each rectangle collection. These two layout patterns at different levels are connected through a \operator{BoundsEvaluator} and its output \variable{Bounds} variable. The \variable{Bounds} of the rectangle marks are not required as an input to \operator{LayoutManager} L2, but they are essential in ensuring the propagation of any changes in the upstream layout. The link between the mark \variable{Bounds} and \operator{LayoutManager} L2 is thus a dependency link. Finally, we introduce two more \operator{BoundsEvaluators} to update the \variable{Bounds} of each rectangle collection and the \variable{Bounds} of the top-level collection whenever any variable involved in L1 or L2 changes its value. 

\begin{figure}[ht]
 \centering \includegraphics[width=1.0\linewidth]{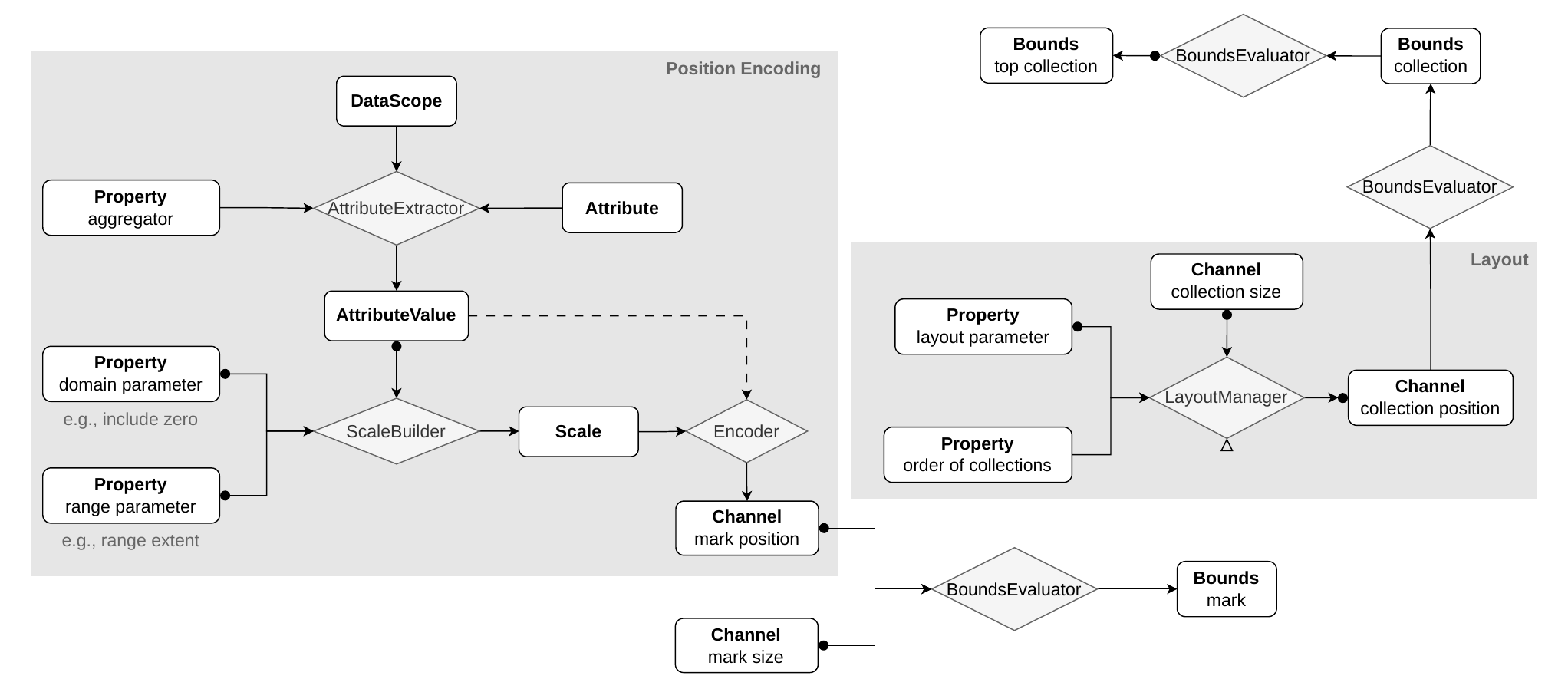}
 \caption{The Encoding-In-Layout Pattern.   \label{fig:encoding-in-layout}}
\end{figure}

The small multiples of calendar heatmaps in Figure \ref{fig:encoding-in-layout} also have a nested structure: within each of the heatmap, we have \enc{x encoding} (\ie the x-position of the rectangle marks encodes \attr{week~number}) and \enc{y encoding} (\ie the y-position encodes \attr{day~of~the~week}). The five heatmaps are then placed in a single-column \layout{grid layout}. Figure \ref{fig:encoding-in-layout} shows the pattern for such structures with \textbf{encodings inside a layout}. The \enc{position encoding} subgraph is consistent with the pattern in Figure \ref{fig:teaser}(b), and the \layout{layout} subgraph is consistent with the pattern in Figure \ref{fig:layout}. Similar to Figure \ref{fig:nested-layouts}, the two subgraphs are connected through a \operator{BoundsEvaluator} and a \variable{Bounds} variable, involving a dependency link.

\section{Code for E3: Stateful Interaction}
Listing \ref{code:e3} shows the code implementing E3 (Section \ref{sec:abstraction}). Lines 1-14 create the static scatter plot. Lines 16-21 specify that clicking a circle mark updates the property ``pinned'' in the state context. Lines 23-35 specify that a change in the state variable ``pinned'' triggers a conditional encoding, where an evaluator checks each circle mark to see if the mark is pinned. The stroke width of each circle is updated accordingly. Line 37-58 specify the behavior and properties of the arrow mark in response to hovering events with the state variable ``pinned'' taken into consideration. Finally, lines 60-77 specify the behavior and properties of the rich text tooltip in response to hovering events with the state variable ``pinned'' taken into consideration.

\begin{lstlisting}[style=js,caption={Specifying the interactive behaviors in E3 (Section \ref{sec:abstraction}): programmers only need to specify the visualization and interaction components, they do not need to understand and construct the dependency graphs.},label={code:e3}]
let scn = msc.scene();
let dt = await msc.csv("datasets/csv/gdp-lifeExp.csv");
let circle = scn.mark("circle", {radius: 6, x: 100, y: 80});
let collection = msc.repeat(circle, dt, {attribute: "Country"});

let xEnc = msc.encode(circle, {attribute:"GDP/PC", channel:"x"});
msc.encode(circle, {attribute:"life expectancy", channel:"y"});
msc.encode(circle, {attribute:"continent", channel:"fillColor"});

scn.axis("x", "GDP/PC", {labelFormat:".2s"});
scn.axis("y", "life expectancy");
scn.legend("fillColor", "continent", {x: 600, y: 250});
scn.gridlines("x", "GDP/PC");
scn.gridlines("y", "life expectancy");
     
let pinTrigger = {event: "click", source: circle},
    pinResponder = {object: scn.state,  properties: ["pinned"]};
let pinUpdater = (evalResult, evtCtx, stateCtx, respObj) => {
    respObj.set("pinned", evtCtx.get("element"));
};
msc.activate(pinTrigger, pinResponder, undefined, pinUpdater);

let stTrigger = {event:"change", source:scn.state.var("pinned")},
    cirResponder = {object: circle, properties: ["strokeWidth"]};
let evaluator1 = function(evtCtx, stateCtx, respObj) {
    return respObj === stateCtx.get("pinned");
};
let cirUpdater = (evalResult, evtCtx, stateCtx, respObj) => {
    if (evalResult) {
        respObj.strokeWidth = 3;
    }
};
msc.activate(stTrigger, cirResponder, evaluator1, cirUpdater);

let arrow = scn.mark("arrow", { visibility: "hidden", strokeColor: "#90CAF9", strokeWidth: 2, endSize: 16 });

let hoverTrigger = { event: "hover", source: circle };
let arrResponder = {
    object: arrow,
    properties: ["visibility", "x", "y", "width", "height"]
};
let evaluator2 = function(evtCtx, stateCtx, respCompnt) {
    return evtCtx.get("element") !== undefined && 
    stateCtx.get("pinned") !== evtCtx.get("element");
}
let arrUpdater = (evalResult, evtCtx, stateCtx, respObj) => {
    let pinned = stateCtx.get("pinned"), hovered = evtCtx.get("element");
    if (pinned && evalResult) {
        respObj.x1 = pinned.bounds.x;
        respObj.y1 = pinned.bounds.y;
        respObj.x2 = hovered.bounds.x;
        respObj.y2 = hovered.bounds.y;
        respObj.visibility = "visible";
    } else {
        respObj.visibility = "hidden";
    }
};
msc.activate(hoverTrigger, arrResponder, evaluator2, arrUpdater);

let tooltip = scn.mark("richText", { visibility: "hidden", anchor: ["left", "top"], backgroundColor: "#fff"});
const formatter = new Intl.NumberFormat('en-US', { signDisplay: 'exceptZero' });
let ttResponder = {
    object: tooltip,
    properties: ["visibility", "x", "y", "text"] 
};
let ttUpdater = (evalResult, evtCtx, stateCtx, respObj) => {
    if (evalResult && stateCtx.get("pinned")) {
        let pinned = stateCtx.get("pinned").datum, hovered = evtCtx.get("element").datum;
        respObj.visibility = "visible";
        respObj.x = evtCtx.get("x") + 2;
        respObj.y = evtCtx.get("y") + 2;
        respObj.text = "Comparing <b>" + hovered['Country'] + "</b> to <b>" + pinned['Country'] + "</b>:<br>Life expectancy: " + formatter.format(hovered['Life expectancy'] - pinned['Life expectancy']) + " years<br>GDP per capita: " + formatter.format(hovered['GDP per capita'] - pinned['GDP per capita']);
    } else {
        respObj.visibility = "hidden";
    }
};
msc.activate(hoverTrigger, ttResponder, evaluator2, ttUpdater);

msc.renderer("svg", "svgElement").render(scn);
\end{lstlisting}

\section{User Study Survey Results}
\label{appendix:study_results}
\begin{figure}[ht]
 \centering \includegraphics[width=1.0\linewidth]{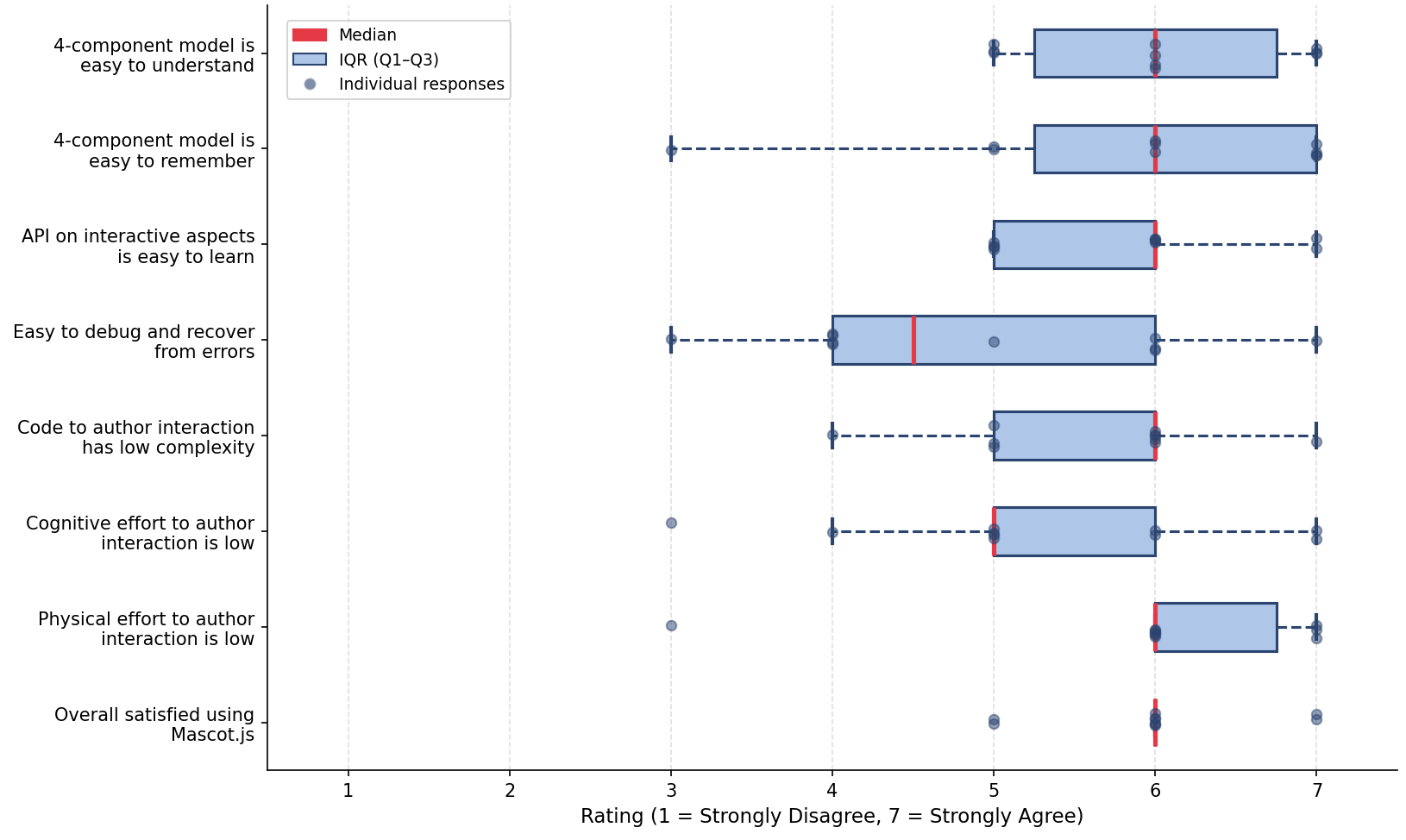}
 \caption{Post-study survey responses.   \label{fig:study_responses}}
\end{figure}

As shown in Figure \ref{fig:study_responses}, participants responded positively across most dimensions of Mascot.js, with the majority of questions receiving median ratings of 6 out of 7. The highest and most consistent ratings were for understanding the 4-component model, physical effort, and overall satisfaction, where responses clustered tightly between 5 and 7, suggesting broad agreement. Learnability of the API and low code complexity were similarly well-received, though with slightly more spread. The clearest pain point was debugging and error recovery, which had the lowest median and the widest distribution, indicating that participants varied considerably in how easy it was for them to recover from mistakes. Cognitive effort was also rated more modestly than other dimensions. The data suggest that Mascot.js is generally perceived as approachable and satisfying to use, but that debugging support warrants further attention.

\begin{figure}[t]
 \centering \includegraphics[width=1.0\linewidth]{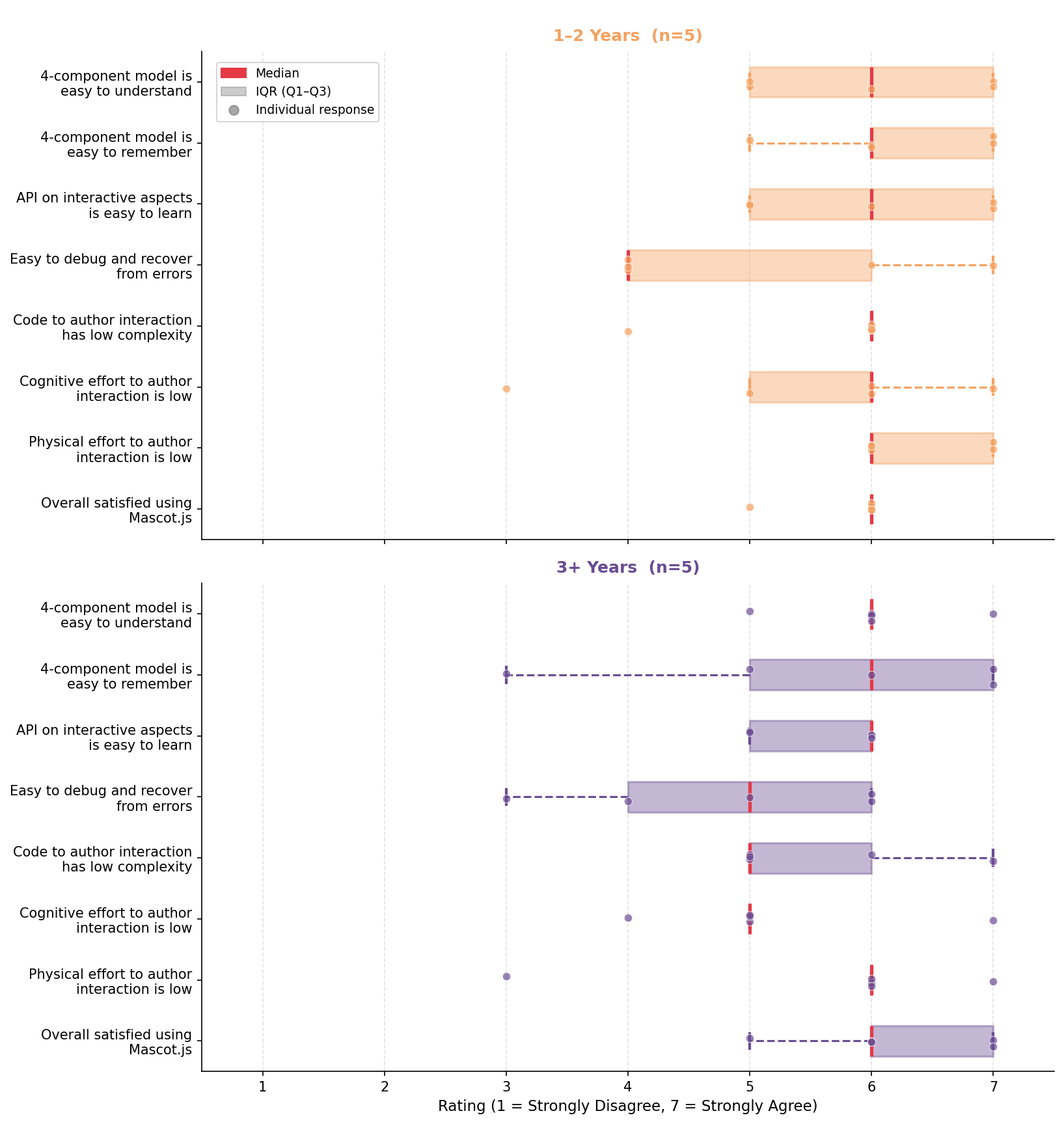}
 \caption{Post-study survey responses by experience.   \label{fig:experience_rating}}
\end{figure}

Figure \ref{fig:experience_rating} shows survey responses broken down by participants' experiences with Vega-Lite/D3 (1-2 years vs. 3+ years). Across most questions, the two experience groups show broadly similar rating patterns, with medians around 6 out of 7 for the majority of dimensions. Both groups identify debugging and error recovery as the weakest area, with lower and more spread-out ratings compared to other questions. The 1–2 years group tends to give slightly more consistent responses with tighter distributions, particularly for physical effort, where their ratings cluster between 6 and 7. The 3+ years group, by contrast, shows more variability across several questions — notably for ease of remembering the 4-component model and physical effort, where a single low rating pulls the whisker down considerably. Interestingly, the 3+ years group rates overall satisfaction marginally higher, suggesting that more experienced users may find greater value in the tool despite having a wider range of opinions on individual aspects. The similarity in medians across groups overall suggests that Mascot.js is perceived comparably regardless of prior experience, though the greater spread in the 3+ years group may reflect higher or more differentiated expectations from users who are already deeply familiar with visualization development.

